\documentclass{IEEEtran}
\usepackage{cite}
\usepackage{amsmath,amssymb,amsfonts}
\usepackage{algorithmic}
\usepackage{graphicx,color}
\usepackage{textcomp}
\usepackage{gensymb}
\usepackage{placeins}
\usepackage[hyphens]{url}
\usepackage[caption=false,font=normalsize,labelfont=sf,textfont=sf]{subfig}
\def\BibTeX{{\rm B\kern-.05em{\sc i\kern-.025em b}\kern-.08em
    T\kern-.1667em\lower.7ex\hbox{E}\kern-.125emX}}
\begin{document}

\title{A Two-Element Capsule Antenna for Near-Field Polarization MIMO in Wireless Capsule Endoscopy}

\author{Pyry Kiviharju, \IEEEmembership{Graduate Student Member, IEEE}, Lauri Vähä-Savo, \IEEEmembership{Member, IEEE}, Clemens Icheln, Kenichi Takizawa, \IEEEmembership{Member, IEEE}, Katsuyuki Haneda, \IEEEmembership{Member, IEEE}
\thanks{This work was supported by the Research Council of Finland, decisions \#338786/WEIRD, \#362750/DISCAS, and \#371367/TORI.}
\thanks{Pyry Kiviharju, Clemens Icheln and Katsuyuki Haneda are with the Department of Electronics and Nanoengineering,
School of Electrical Engineering, Aalto University, 00076 Espoo, Finland
(e-mail: pyry.kiviharju@aalto.fi).}
\thanks{Lauri Vähä-Savo was with the Department of Electronics and Nanoengineering, School of Electrical Engineering, Aalto University, 00076 Espoo, Finland. He is now with Nokia Bell Labs Finland, Espoo, Finland.}
\thanks{Kenichi Takizawa is with the National Institute of Information and Communications Technology, Sendai, Japan.}}
\maketitle

\begin{abstract}
Multiple-Input-Multiple-Output (MIMO) technology is considered to be a viable method for spectral efficiency and data rate enhancement in wireless communications. This article presents a two-element capsule antenna with orthogonal current distributions for polarization multiplexing in a wireless capsule endoscopy (WCE) application at $915$~MHz. The proposed antenna was evaluated in a MIMO link where it served as the transmitter and an existing on-body antenna was the receiver. The performance was assessed in two distinct environments: an anatomically accurate human body model using full-wave (FW) simulations, and a homogeneous liquid tissue phantom, using both FW simulations and measurements. The former was used to evaluate the performance in a realistic environment and to demonstrate the superiority of the proposed capsule antenna against existing two-element antennas in the literature, while the latter aimed at verifying the antenna design by comparison of simulated and measured MIMO links. We studied channel capacity and eigenvalues for the comparisons. The MIMO link with the proposed antenna provided nearly doubled median channel capacity relative to a baseline single-input-single-output link, consistently in the simulations and measurements. Our results prove the suitability of the proposed antenna for polarization multiplexing and, in terms of channel capacity, establish it as the best conformal two-element capsule antenna design at $915$~MHz.
\end{abstract}


\maketitle

\section{INTRODUCTION}
\IEEEPARstart{W}{ireless} capsule endoscopy (WCE) has become the standard method for human small intestine disease examination~\cite{Iddan2000}. In contrast to a rigid endoscope, a WCE can access the winding small intestine and is less invasive, providing a more comfortable experience for patients. Since the inception of the technology in 2000, WCEs have been successful in diagnosing various human gastrointestinal (GI) tract disorders, such as Crohn's disease and tumors~\cite{Iddan2000, Meron2000, Cobrin2006, Hara2006, Ge2004}.

For accurate diagnostics of disorders, the images transmitted by a WCE should have a high resolution and frame rate. The required data rates pose challenges for the wireless data telemetry link from inside to outside of the human body. Over the past few years, multiple-input-multiple-output (MIMO) technology has attracted the attention of researchers for enhancing data rates without increasing the frequency band usage. In cellular mobile scenarios, MIMO enhances the channel capacity mainly by taking advantage of multipath propagation. Consequently, multi-element antenna designs for the WCE scenario have also been studied. Antennas having two elements can be found in~\cite{Alazemi2022, Iqbal2024a, Chang2024, Faisal2024, MisbahUnNoor2025, Kiani2025, Li2025, Sun2025} and references therein, whereas up to four elements have been considered~\cite{Wang2022b, Shah2023, Shah2024, Iqbal2024, Abbas2025, Qamar2026}. However, when evaluating MIMO links in the WCE scenario, these studies implicitly assume rich multipath and far-field wave propagation as in the cellular scenarios~\cite{Li2012}, except~\cite{Kiani2025}, which does not assume rich multipath but still far-field conditions. This assumption is most likely invalid in the WCE scenario due to the short link distance inside a human body. Rather, the propagation condition of a WCE MIMO link is similar to that of an underwater line-of-sight (LOS) MIMO channel~\cite{Takizawa2021, Takizawa2021_EuCNC}. Moreover, a WCE link relies on near-fields of the antennas, for which radiation patterns cannot be uniquely defined.

One option to take advantage of MIMO without relying on multipaths is the use of different polarizations. In this article, the term polarization refers to the electric field spatial orientation, regardless of the field zone. It has been reported that multi-polarized antennas in free-space LOS-{\it only} MIMO links leverage up to two effective communication channels, leading to the diversity or multiplexing gain~\cite[Section IV-C]{Jensen2004},~\cite{Svantesson2004}. Multiple-port antennas exploiting orthogonal polarizations, and equivalently those exciting orthogonal current distributions, can be designed with a smaller form factor than antenna arrays, making them suitable for implementation on the capsule side. Despite this apparent advantage, to the best of the authors' knowledge, no study except the present authors' own~\cite{Kiviharju2024} has addressed the polarization MIMO in WCE context. The previous work~\cite{Kiviharju2024} utilized idealistic small loop antennas with perfect matching and isolation at link ends, lacking a practical design of a multiple-port antenna exciting orthogonal current distributions.

To expand our previous study~\cite{Kiviharju2024}, this paper proposes a novel conformal two-element WCE antenna for polarization multiplexing operating in the $915$~MHz industrial-scientific-medical (ISM) band~\cite{ITUreg}, spanning from $902$~MHz to $928$~MHz. The fitness of the proposed antenna was evaluated by pairing it with a state-of-the-art dual-polarized on-body antenna~\cite{Wang2022c} and computing the capacity and eigenvalues of the corresponding MIMO channel. The conventional far-field MIMO performance metrics, such as envelope correlation coefficient and mean effective gain, were not used for benchmarking the antenna since they are not valid in near-field conditions. Instead, following our methodology to analyze near-field radio channels for single-antenna WCE radio links~\cite{Miah19_TAP}, we derived our MIMO radio channel responses from full-wave (FW) simulations and experimental measurements for their validation. With this methodology, we effectively accounted for the near-field radiation of antennas, which leads to a radio channel dominated by direct electromagnetic coupling.

The MIMO link was evaluated in two environments as follows: by FW simulations in a realistic three-dimensional (3D) body model and by FW simulations and measurements in a muscle tissue phantom. Our designed antenna improved the median channel capacity compared to a single-antenna channel by a factor of $1.70$ to $1.96$ in both environments and yielded the highest channel capacity among the existing conformal two-element antennas operating in the $915$~MHz ISM band. In addition, simulated and measured channel capacity and eigenvalues in a phantom environment showed agreement. These results prove the suitability of the proposed antenna for polarization multiplexing and demonstrate the efficacy of our scientific approach and implementation.

The remainder of this paper is organized as follows. Section~\ref{sec:antdesign} presents the proposed two-element capsule antenna design and its operating principles. Section~\ref{sec:MIMOsim} describes the FW simulation setup of a MIMO link using a 3D human body model and compares the radio link performance of the proposed antenna against existing conformal two-element capsule antennas at $915$~MHz. Section~\ref{sec:MIMOmeas} shows the MIMO link evaluation in the phantom scenario and compares the simulations with measurements.  Finally, Section~\ref{conclusion} concludes the paper. 

\section{Two-Element Capsule Antenna Design}\label{sec:antdesign}
\begin{figure}[h]
    \centering
    \hspace{-20pt}
    \subfloat[\label{fig:ant_top}]{\includegraphics[height=5cm]{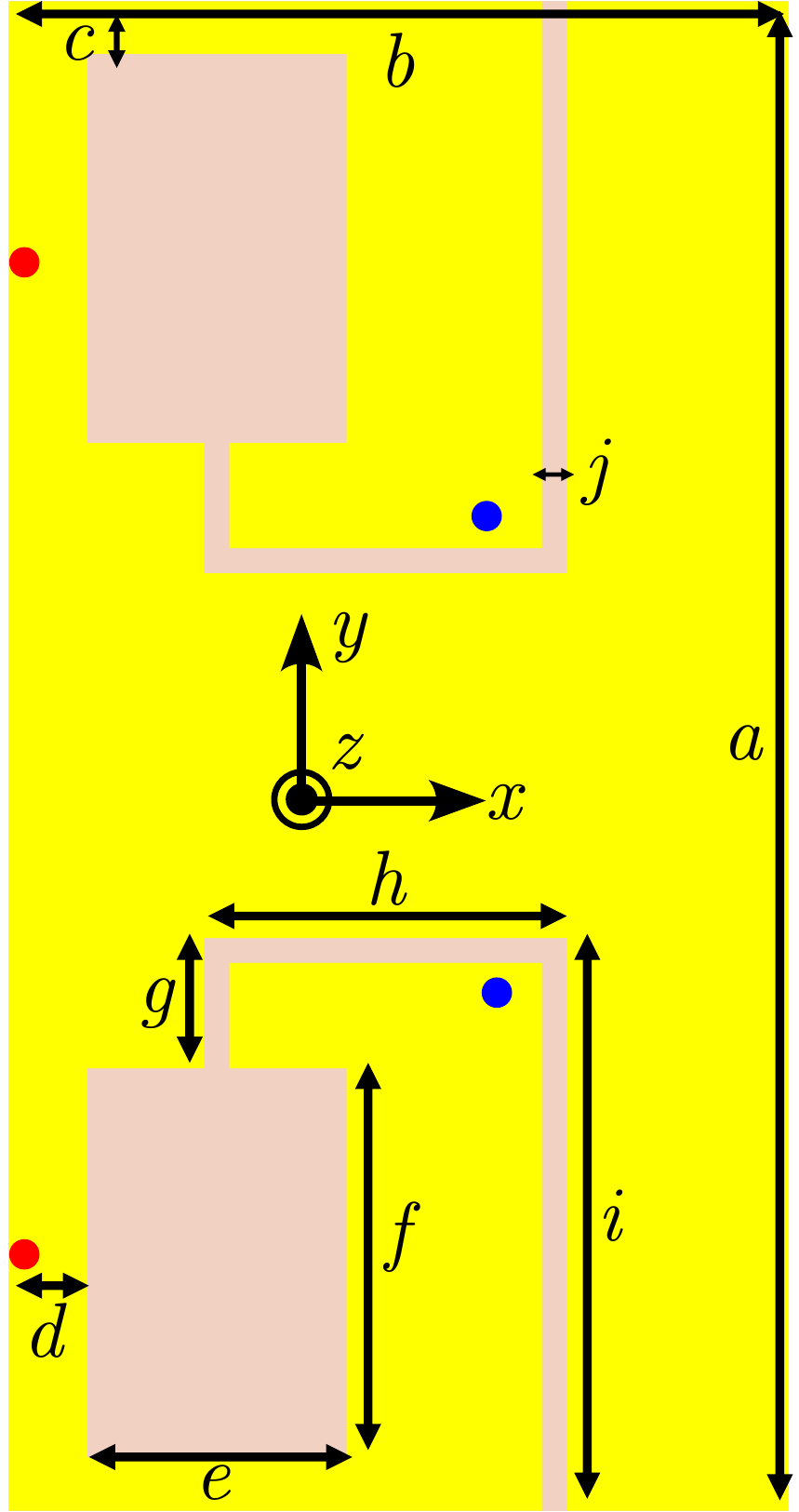}}
    \hspace{30pt}
    \subfloat[\label{fig:antbottom}]{\includegraphics[height=5cm]{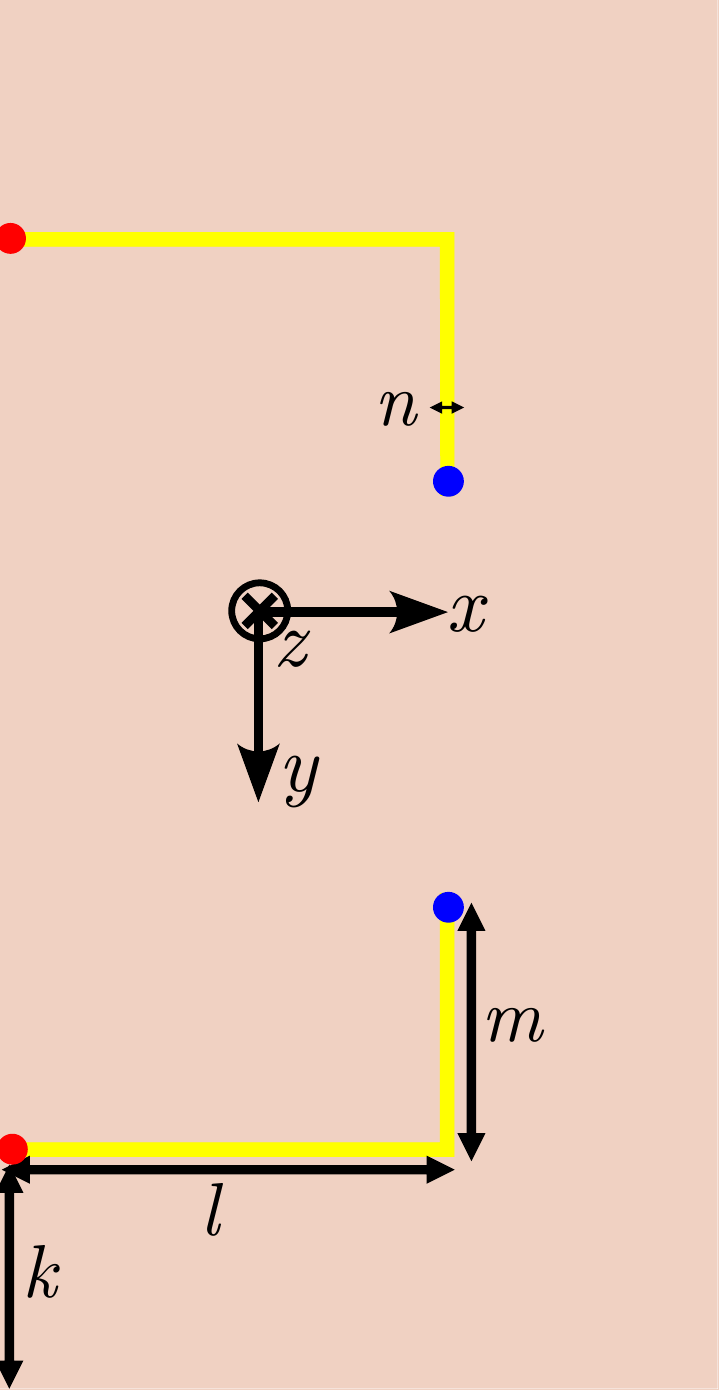}}
    \hfill
    \subfloat[\label{fig:antwrapped}]{\includegraphics[width=0.7\columnwidth]{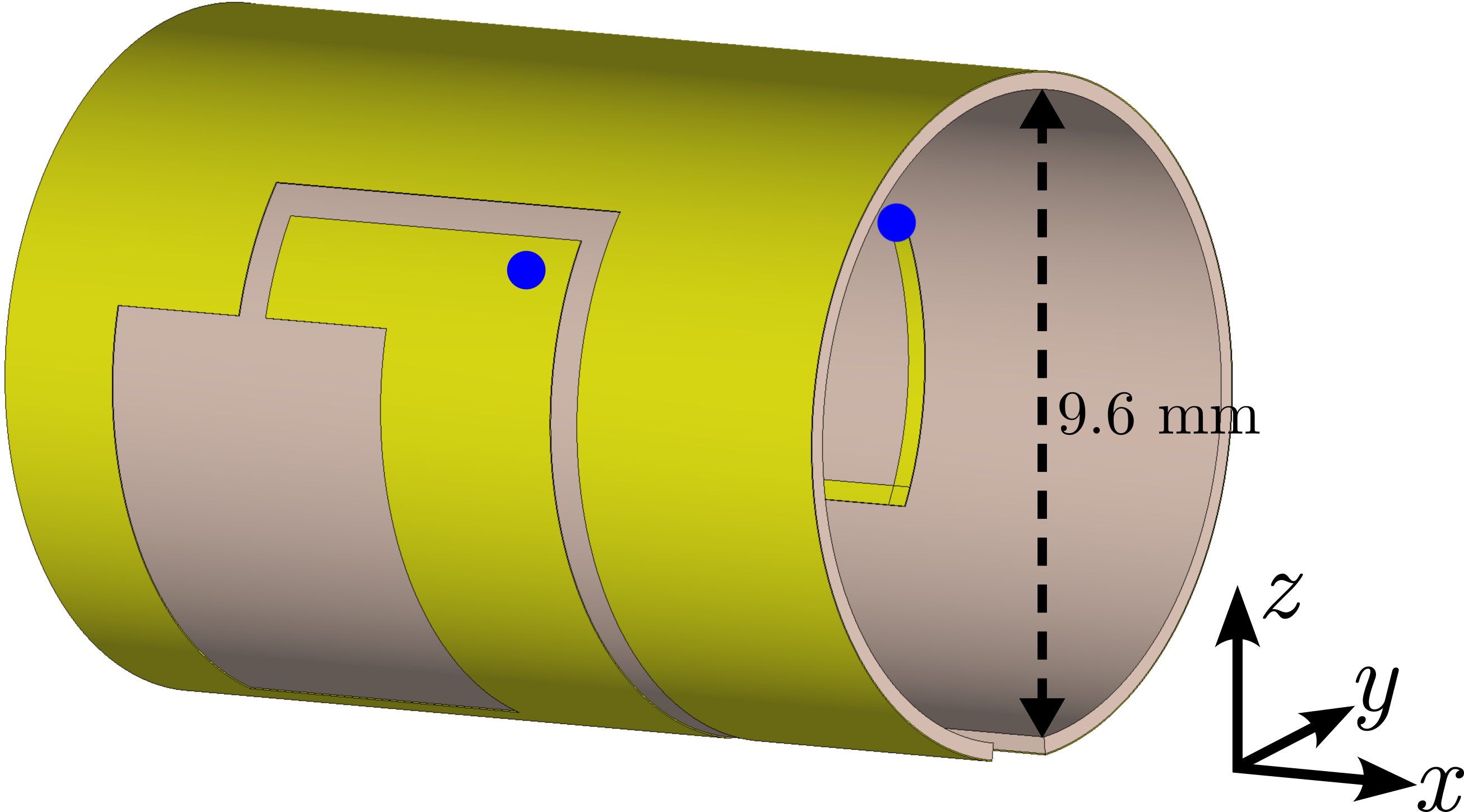}}
    \caption{Proposed two-element WCE capsule antenna a) PCB front/outer side b) PCB back/inner side and c) conformed into cylindrical form. The yellow and light pink colors represent copper and dielectric substrate, respectively. The red dots indicate the feed positions, while the blue dots mark locations of vias.}
    \label{fig:steppedslot}
\end{figure}

Our capsule antenna is a conformal two-element slot antenna operating at the $915$~MHz ISM band, designed on a two-layer printed circuit board (PCB). Fig.~\ref{fig:steppedslot} shows the antenna PCB before and after conforming to the inner wall of a capsule. The conformation involves bending so that the short sides of the planar substrate face each other, leaving a $1.3$~mm gap between the sides. The two elements are mirror images of each other. The dimensions of the antenna are provided in Table~\ref{tab:antennadim}.
The material of the substrate is RO3010 from Rogers Corporation~\cite{RogersCorp}, with $\varepsilon_{\rm r} = 10.2$ and $\tan \delta = 0.002$. As RO3010 is not a flexible material, a PCB thickness of $0.25$~mm and a copper thickness of $18\ \mu \mathrm{m}$ were opted to facilitate the conforming of the antenna. 
\begin{table}[h]
    \centering
    \caption{Optimized dimensions of the proposed antenna (mm)}
    \begin{tabular}{|cc|cc|cc|cc|cc|cc|cc|}
    \hline
    a & 29.0 & d & 1.0 & g & 1.6 & j & 0.5 & m & 5.5\\
    \hline
     b & 15.0 & e & 5.0 & h & 6.7 & k & 4.0 & n & 0.3\\
     \hline
     c & 1.0 & f & 8.0 & i & 11.5 & l & 9.5 & $-$ & $-$\\
     \hline
    \end{tabular}
    \label{tab:antennadim}
\end{table}

\subsection{Antenna Operation Principle and Design Evolution}
A WCE capsule antenna must be small enough to fit in the extremely limited real estate of a $26$-by-$11$~mm$^2$ standard capsule. Thus, one suitable basis design is a quarter-wave slot, which was selected as the starting point for the antenna design. Details of this antenna type are found in, e.g.,~\cite{Balanis2005},\cite{Stutzman2013}.
\begin{figure}[h]
    \centering
    \includegraphics[width=0.9\columnwidth]{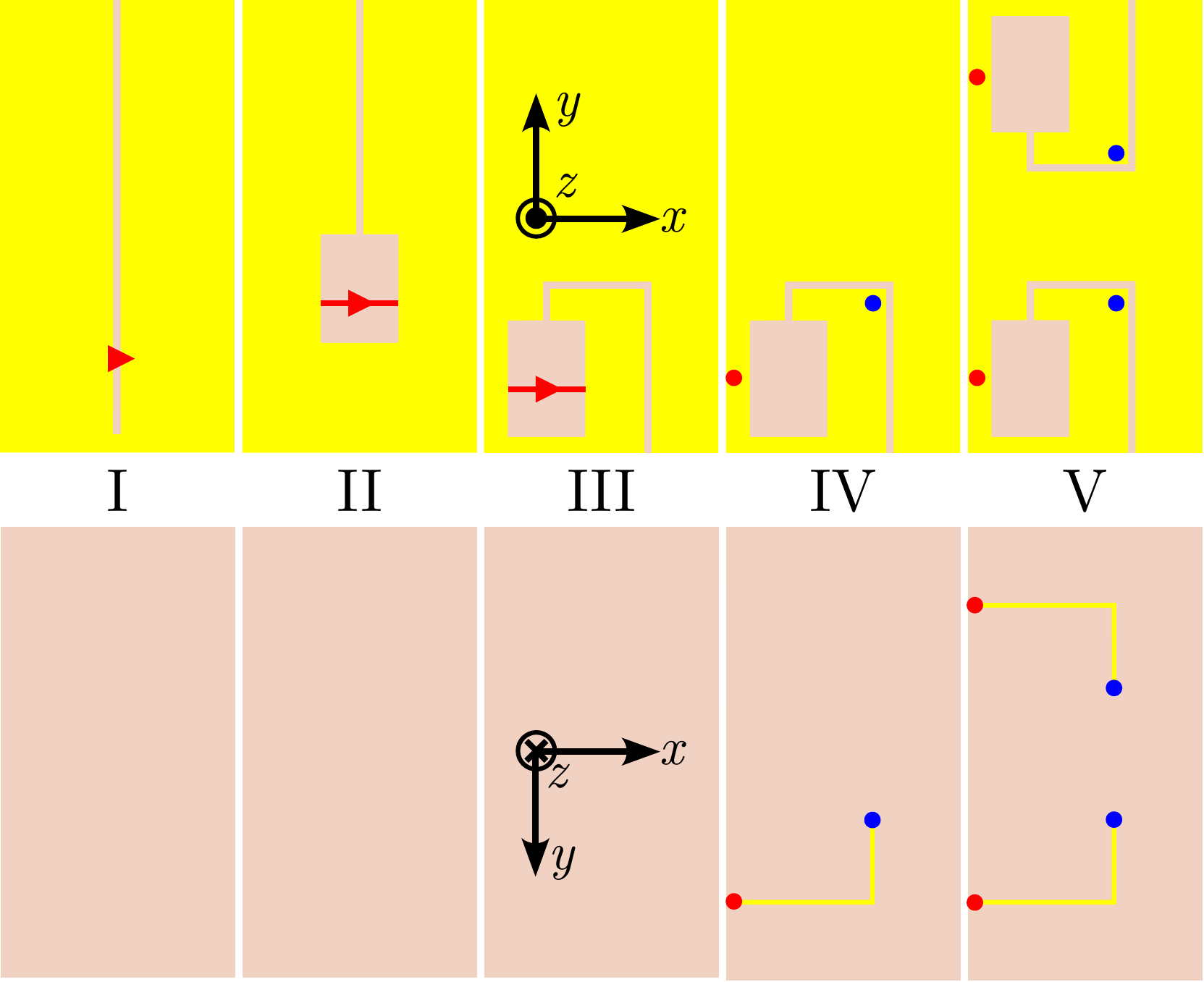}
    \caption{Design phases of the proposed two-element stepped impedance slot antenna. The upper row shows the front side of the PCB, and the lower row shows the back side. The yellow and light pink colors represent copper and dielectric substrate, respectively. Feed positions are indicated with red color, and via positions are shown with blue dots.}
    \label{fig:designevolution}
\end{figure}
The design progressed in six phases, with phases~I to~V illustrated in Fig.~\ref{fig:designevolution}. In each phase, the antenna was simulated while enclosed in a spherical intestine tissue phantom with a radius of $50$~mm. The size of the phantom was found sufficient, as the reflection coefficient was the same as in a larger $100$~mm phantom. Implanted antennas are typically covered with a dielectric insulation to prevent current leakage into tissues, which exacerbates the radiation efficiency. We surrounded the antenna with a cuboidal air gap extending $0.5$~mm in each direction from the antenna PCB. This thickness resembles a realistic shell thickness of an endoscopic capsule.

First in phase~I, a quarter-wave slot resonating at $915$~MHz was cut on the top metal of a single-layer PCB. However, to facilitate placing another slot on the same PCB, two different size reduction methods, i.e., impedance step~\cite{Sagawa1997} and meandering, were applied. Adding an adjacent parallel slot to phase~I was not feasible as two collinear slots on a curved PCB couple tightly, as shown later in the present section. The impedance step has been utilized previously in implantable antenna designs~\cite{Nikolayev2017} to miniaturize a microstrip resonator antenna, where a $\lambda/2$ self-resonant antenna was created by two low-impedance and one high-impedance sections. A $\lambda/4$ version of the similar antenna~\cite{Sagawa1997} consisted of one low-impedance section connected to a high-impedance section, which was shorted to the ground. The impedance across the sections changes abruptly, akin to a step rather than gradually. However, since our antenna was a slot rather than a microstrip antenna, the geometry of~\cite{Sagawa1997} could not be directly employed. For slotlines, the impedance increases with the slot width~\cite{Janaswamy1986, Lee1991}, and hence the high-impedance section had to be wider than the low-impedance section. Therefore, the first stage of miniaturization was achieved by widening the slot on the feed port side. The width ratio between the low and high-impedance sections needs to be as large as possible to facilitate miniaturization~\cite{Sagawa1997}. The antenna shape with the impedance step is shown in phase~II.

The antenna was further miniaturized by meandering the narrow low-impedance section. Two $90\degree$ bends were sufficient for a resonance at $915$~MHz. To make a space for the second antenna, the high-impedance section was moved to the negative $x$ and $y$ directions, resulting in the phase~III design.

Up to this stage, a balanced discrete port across the high-impedance slot section was used to excite the antenna. Since the antenna is in practice fed by a connector via an unbalanced coaxial cable, microstrip feeding~\cite{Latif2005} was considered here. It provides sufficient bandwidth and serves as a balanced-to-unbalanced transformer (balun). Hence, an L-shaped microstrip feed for the antenna was created on the opposite side of the PCB. A discrete excitation port was placed between the feed line and the front metal at the PCB edge. In phase~IV of Fig.~\ref{fig:designevolution}, the discrete port position is indicated by the red dot. Currents on the horizontal feed section on the back side couple to the wide slot section on the front side, exciting a magnetic current. The horizontal feed section was positioned approximately at the same $y$-coordinate as the discrete feed in phases~II and III. For tuning the resonance to the band of interest, a short-circuiting via was placed at the end of the feed line, indicated by the blue dot in Fig.~\ref{fig:designevolution}, phase~IV.

As phase~V, the antenna was copied and mirrored with respect to the $x$-axis to create a two-element antenna, while the sixth and final phase was conforming the antenna into a cylindrical shape, as depicted in Fig.~\ref{fig:antwrapped}. In this format, the antenna was insulated by placing an 0.4~mm thick sheet of polylactic acid (PLA) on its outer surface. For PLA, $\varepsilon_{\rm r} = 2.7$ was assumed~\cite{Felicio2016, UltimakerMaterials}. This material and thickness were selected to be compliant with manufacturing by 3D-printing. 

\begin{figure}[h]
    \centering
     \subfloat[\label{fig:coupling_par}]{\includegraphics[height=3.5cm]{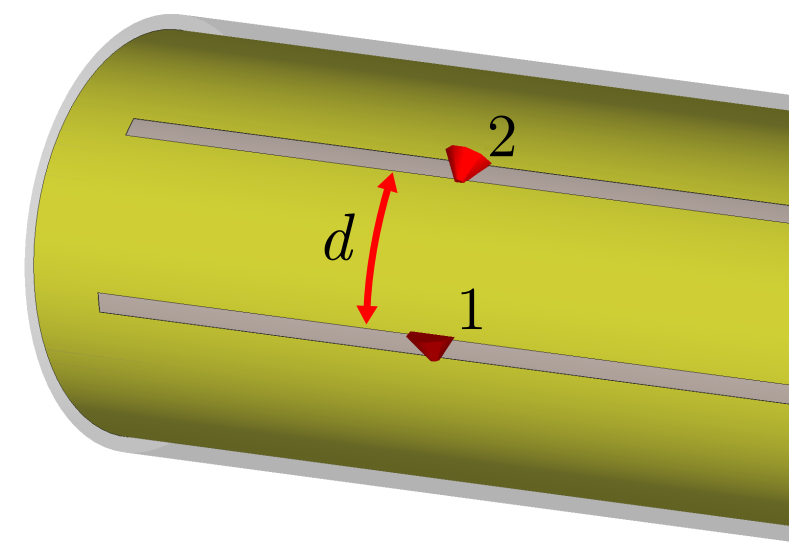}}
     \hfill
    \subfloat[\label{fig:coupling_collinear}]{\includegraphics[height=3.2cm]{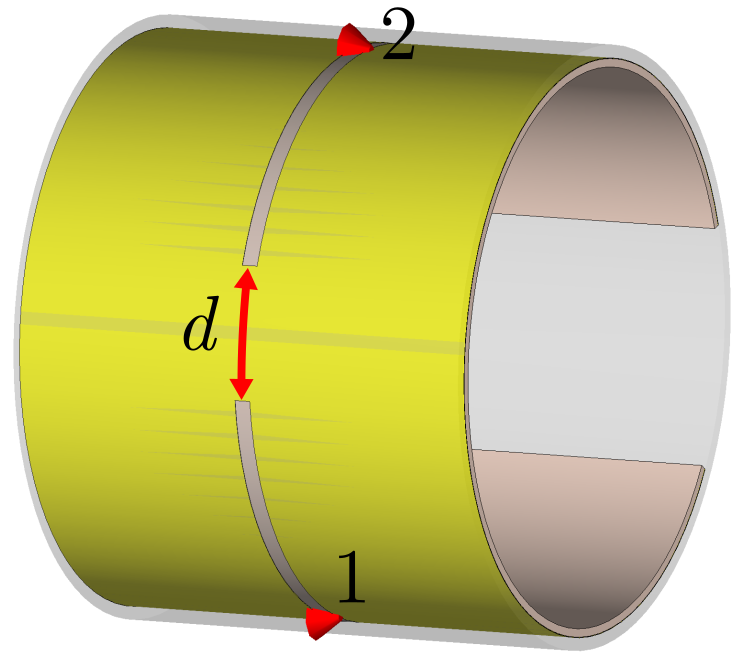}}
    \caption{Two slots on a curved PCB a) parallel b) collinear. The red cones indicate feed port locations. In a), the slots extend to the right beyond the image border.}
    \label{fig:coupling_sim}
\end{figure}

\begin{figure}[h]
    \centering
    \includegraphics[width=\columnwidth]{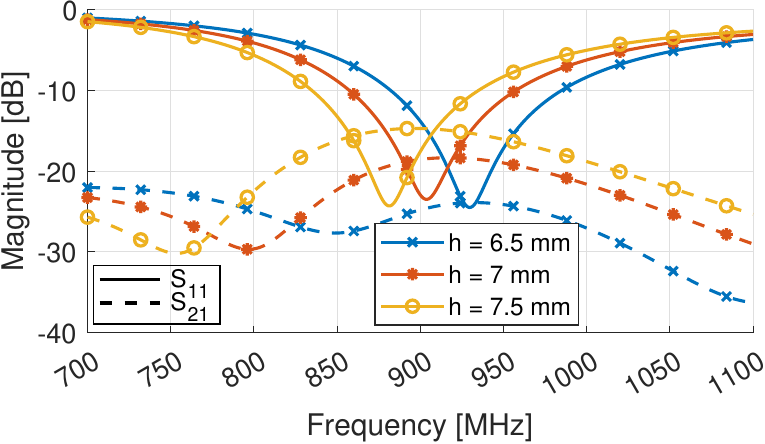}
    \caption{Matching ($|S_{11}|$) and coupling ($|S_{21}|$) levels as a function of slot length $h$ in Fig.~\ref{fig:ant_top}.}
    \label{fig:coupling_vs_h}
\end{figure}

In addition to a proper input matching in terms of the scattering parameter $\lvert S_{11} \rvert < -10$~dB across the band of interest, for MIMO operation, a two-element antenna should also maintain a coupling level of $\lvert S_{21} \rvert < -20$~dB between the antenna ports~\cite{Mei2018}. The coupling between the ports of the proposed antenna can be analyzed by studying the coupling of open-ended slot antennas that are either parallel or collinear to each other on a curved PCB, as illustrated in Fig.~\ref{fig:coupling_sim}. These setups were simulated with CST Studio Suite~\cite{CST} inside a homogeneous intestine tissue phantom while the PCBs were insulated with a $0.4$~mm PLA layer. The slot distances $d$ were varied from $4$~mm to $10$~mm, and $\lvert S_{21} \rvert$ was observed at the resonant frequency of the slots.
For parallel slots, the coupling between the ports ranged from $-5$~dB to $-10$~dB. On the other hand, for the collinear slots, the coupling ranged from $-17$~dB to $-19$~dB. This indicated that the parallel slots couple tightly while the collinear slots do not. Hence, the coupling between the ports of the proposed antenna is mostly attributed to the parallel slot sections whose dimension is labeled as $h$ in Fig.~\ref{fig:ant_top}. The extent of coupling depends on the length $h$ as plotted in Fig.~\ref{fig:coupling_vs_h}, showing that the coupling increases rapidly as $h$ is increased.
Hence, decoupling was achieved by keeping $h$ as small as possible, although it cannot be made arbitrarily small because shortening it increases the resonant frequency. The optimized dimensions of the proposed antenna mentioned earlier in Table~\ref{tab:antennadim} meet $\lvert S_{11} \rvert < -10$~dB and $\lvert S_{21} \rvert < -20$~dB at $915$~MHz for $1$~MHz bandwidth inside a homogeneous intestine tissue phantom.

\subsection{Characteristics of the Antenna}
\begin{figure}[h]
    \centering
    \includegraphics[width=\columnwidth]{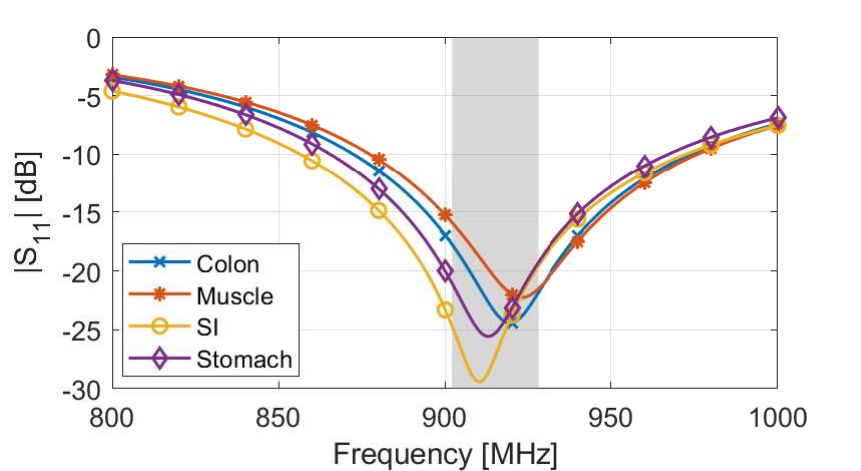}
    \caption{Reflection coefficient magnitude $\lvert S_{11}\rvert$ of the proposed antenna in different GI tract tissues (SI = small intestine). The $915$~MHz ISM band is marked with the gray rectangle.}
    \label{fig:sp_in_tissues}
\end{figure}

This subsection characterizes the proposed antenna alone, while link simulations are presented in Section~\ref{sec:MIMOsim}. All simulations were performed with the frequency domain finite element method (FEM) solver of CST Studio Suite~\cite{CST}. A capsule antenna is required to operate inside each tissue in the human GI tract. Table~\ref{tab:tissue_parameters} shows the dielectric properties of human GI tract tissues at $915$~MHz. Figure~\ref{fig:sp_in_tissues} shows $\lvert S_{11}\rvert < -10$~dB in a $50$~mm radius spherical phantom of those tissues across the intended band, proving
\begin{table}[h]
    \centering
    \caption{Real and imaginary parts of tissue permittivity at 915 MHz}
    \begin{tabular}{|c|c|c|c|c|}
    \hline
     Tissue  & Colon &  Muscle & Small intestine & Stomach \\
     \hline
     $\varepsilon_{\rm r}^\prime$    & $57.8$ & $55.0$ & $59.1$ & $64.9$ \\
     \hline
     $\varepsilon_{\rm r}^{\prime\prime}$ & $21.0$ & $18.6$ & $42.4$ & $23.4$ \\
     \hline
    \end{tabular}
    \label{tab:tissue_parameters}
\end{table}
robustness of the proposed antenna against tissue-induced resonance shifts.

\begin{figure}[h]
    \centering
    \includegraphics[width=0.9\columnwidth]{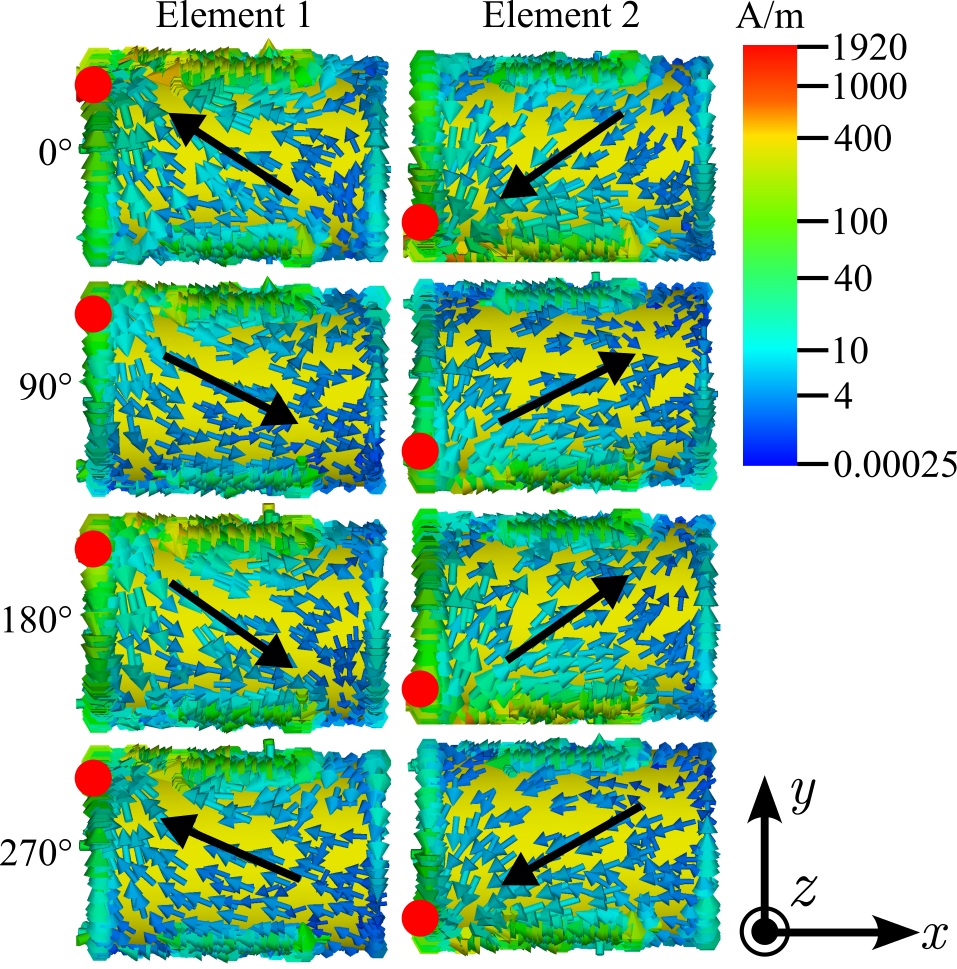}
    \caption{Surface current distribution of the proposed antenna at 915 MHz when conformed as Fig.~\ref{fig:antwrapped}. The first and second columns correspond to elements 1 and 2, respectively, while rows represent different phases of the time-harmonic current. For each case, the main current direction is indicated by the black arrow, and the red dots show the location of the discrete feed ports.
    }
    \label{fig:surfcurrent}
\end{figure}

Figure~\ref{fig:surfcurrent} shows the surface current of the conformed proposed antenna, implanted in a $50$~mm radius intestine tissue phantom. The magnitude is indicated by the color and size of the arrows.
The figure shows the antenna seen from the positive $z$ direction, i.e., the slots are at the top and bottom edges of each subfigure; Fig.~\ref{fig:antwrapped} can be referred to for the coordinate system. Although the current magnitude is higher at the slots than in the central part of the antenna, the slot currents contribute little to the radiation. Rather, the slots act as coupling elements for exciting orthogonal current modes in the center, as indicated by the plot. These current modes then lead to orthogonal near-field radiation and polarization multiplexing capability.

\begin{figure}[th]
    \centering
    \includegraphics[width=\columnwidth]{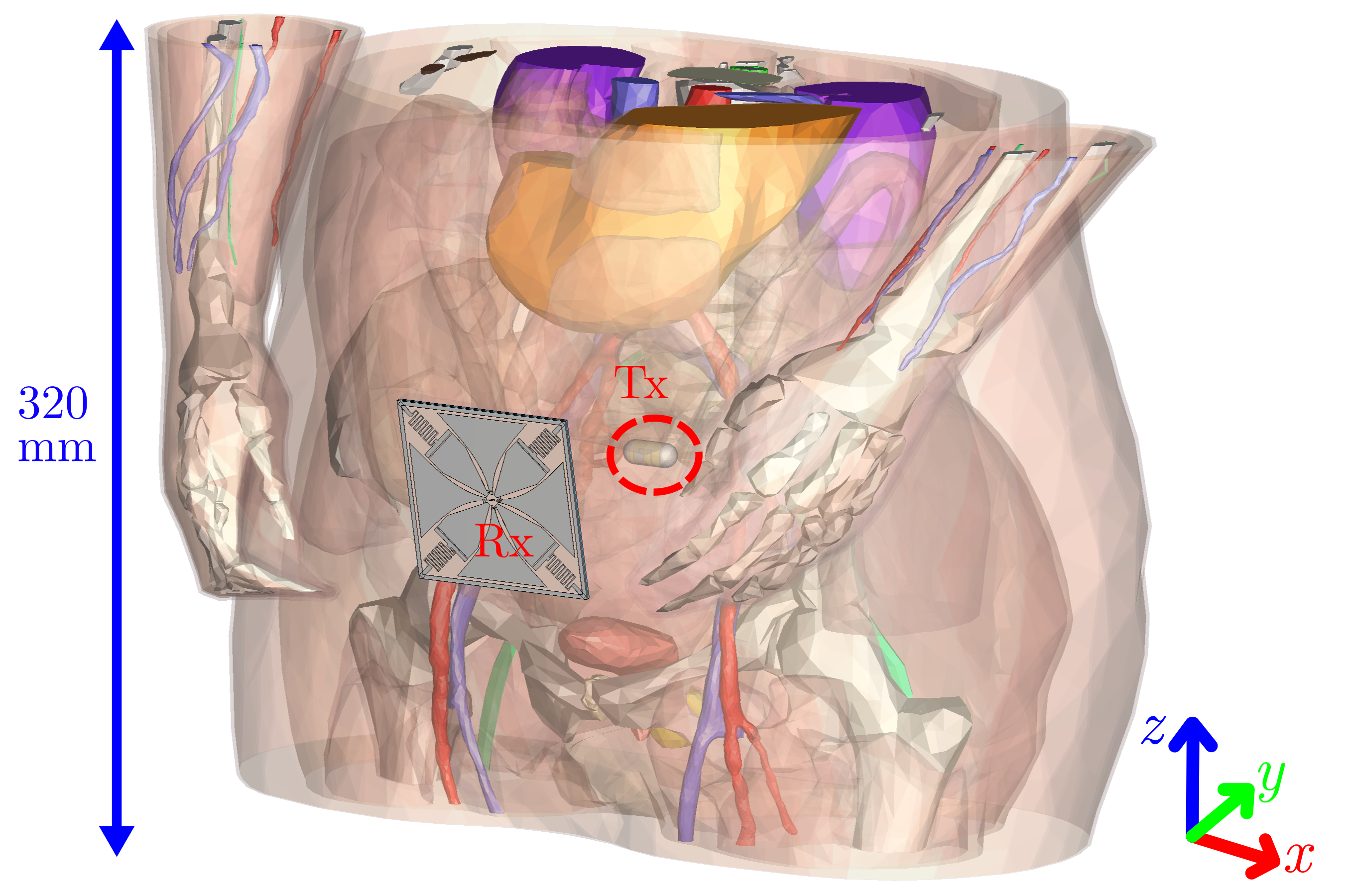}
    \caption{SAR and link simulation setup with "Nelly" 3D human model. Tissues of the model have been made transparent to facilitate the visibility of the capsule Tx antenna.}
    \label{fig:simsetup}
\end{figure}

To ensure the safety of the human body against radio frequency (RF) exposure, a capsule antenna must comply with the specific absorption rate (SAR) limitations. According to the IEEE C95.1-2019 standard, the maximum SAR value averaged over $10$~g of tissue is set to $2$~W/kg~\cite{ICES2019}. The SAR of the proposed antenna was evaluated in the human 3D model detailed in Fig.~\ref{fig:simsetup}. The method to compute SAR for a multi-element antenna depends on the data transmission scheme~\cite{Perentos2012}. For polarization multiplexing, the SAR needs to be evaluated assuming uncorrelated sources, leading to the total exposure being the power sum of exposures due to individual antenna elements. The IEEE 802.15.6 standard for wireless body area networks~\cite{IEEEWban} regulates the maximum input power of an implanted antenna at the $915$~MHz ISM band to be $-10$~dBm. Splitting this power equally across both antenna ports, the maximum SAR was $0.004$~W/kg when the capsule antenna was in the intestine of the model. This low SAR value verifies that the RF exposure level from the proposed antenna is regulation-compliant.

\section{MIMO Link Simulation in a Realistic 3D Body Model}\label{sec:MIMOsim}
\subsection{Preliminaries}\label{preliminaries}
A narrowband MIMO radio channel with $N_{\rm r}$ receive (Rx) and $N_{\rm t}$ transmit (Tx) antennas is characterized by
\begin{equation}\label{eq:mimosignal}
    \mathbf{y} = \mathbf{Hx} + \mathbf{n},
\end{equation}
where $\mathbf{y} \in \mathbb{C}^{N_{\rm r}}$ and $\mathbf{x} \in \mathbb{C}^{N_{\rm t}}$ are the complex Rx and Tx voltage vectors when their characteristic impedance is $50~\Omega$, $\mathbf{H} \in \mathbb{C}^{N_{\rm r}\times N_{\rm t}}$ is the channel matrix, and $\mathbf{n}\in \mathbb{C}^{N_{\rm r}}$ is a complex Gaussian noise vector at the Rx. The channel matrix $\mathbf{H}$ consists of the forward scattering parameters of the radio channel between each Tx and Rx antenna. Our approach to compute $\mathbf{H}$ was through FW simulations of antennas and the human body~\cite{Kiviharju2024}.

The capacity of a MIMO channel is estimated as
\begin{equation}\label{eq:capacity}
    C = B\cdot\log_2 \det \left[ \mathbf{I}_{N_{\rm r}} + \frac{P_{\rm t}}{N_{\rm t} P_{\rm n}}\mathbf{H}\mathbf{H}^{\textnormal{H}}\right] \quad [\textnormal{bits/s}],
\end{equation}
where $\mathbf{I}_{N_{\rm r}}$ is the $N_{\rm r} \times N_{\rm r}$ square identity matrix. In addition, $B$ is the bandwidth, while $P_{\rm t}$ and $P_{\rm n}$ refer to the Tx signal and Rx noise powers, the latter of which is obtained as
\begin{equation}
    P_\mathrm{n} = kTB\cdot NF_{\mathrm{rec}},
\end{equation}
where $k$ is the Boltzmann constant, $T$ is the temperature, and $NF_\mathrm{rec}$ is the receiver noise figure; 
\eqref{eq:capacity} holds when the Rx has no channel state information. An alternative formulation for the channel capacity using eigenvalues of $\mathbf{HH}^\mathrm{H}$ is
\begin{equation}\label{eq:capacity_sigma}
    C = \sum_{i=1}^{\min{(N_{\rm r}, N_{\rm t})}}B\log_2\left(1 + \frac{P_{\mathrm{t}}}{N_{\mathrm{t}}P_{\mathrm{n}}}\lambda_i\right),
\end{equation}
where $\lambda_i$ is the $i$th eigenvalue of $\mathbf{H}\mathbf{H}^\mathrm{H}$;~\eqref{eq:capacity_sigma} shows the capacity of a MIMO channel is the sum of those of eigenchannels, each determined by the magnitude of eigenvalues. The eigenvalues can be used to estimate the channel rank $M_{\mathbf{H}}$. It is the number of eigenvalues falling into the dynamic range with a $6$~dB margin as:
\begin{equation}
    M_{\mathbf{H}} = \sum_{i=1}^{\min{(N_{\rm r}, N_{\rm t})}}  I\{\lambda_i(\mathbf{HH^\mathrm{H}})|_{\mathrm{dB}} > (-DR+ 6)\} , 
\end{equation}
where $DR = 10\log_{10}(P_\mathrm{t}/(N_\mathrm{t}P_\mathrm{n}))$ is the dynamic range in decibels and the function $I\{(\cdot)\}$ returns $1$ if the condition $(\cdot)$ is met, otherwise $0$. In other words, we define an eigenchannel as meaningful if it has an SNR greater than $6$~dB. In case $M_\mathbf{H} = \min(N_\mathrm{r}, N_\mathrm{t})$, the channel is full-rank; otherwise, it is rank-deficient. 

\subsection{Simulation setup}\label{sec:simsetup}
In this section, we evaluated a MIMO radio link realized by the proposed capsule antenna introduced in Section~\ref{sec:antdesign}. It was used as the Tx antenna, while a dual-polarized double-bowtie antenna published in~\cite{Wang2022c} was used as the on-body Rx antenna, thereby $N_{\rm r} = N_{\rm t}=2$. For the remainder of this paper, the terms Tx and Rx will be used to refer to the capsule and on-body antennas, respectively. The simulations of the $\mathbf{H}$-matrix were performed with the FEM solver of CST Studio Suite~\cite{CST}. The simulation setup is shown in Fig.~\ref{fig:simsetup}. To model a realistic operation environment, we used the "Nelly" 3D human model from the CST Bio Models library, which supports the FEM solver.

Two Tx locations, Tx$_\mathrm{a}$ and Tx$_\mathrm{b}$ were considered, corresponding to shallow and deep capsule locations inside the body model, while the on-body Rx antenna remained fixed at a single location. The coordinates of Tx and Rx are provided in Table~\ref{tab:positions}.
\begin{table}[th]
    \caption{Positioning of the Tx and Rx antennas}
    \centering
    \begin{tabular}{|c|c|c|c|}
    \hline
        & $x$ (mm) & $y$ (mm) & $z$ (mm) \\
            \hline
       Tx$_\mathrm{a}$ & $15$ & $40$ & $-613$ \\
        \hline
       Tx$_\mathrm{b}$ & $15$ & $-23$ & $-613$ \\
        \hline
         Rx & $15$ & $-102$ & $-613$ \\
         \hline
\end{tabular}
    \label{tab:positions}
\end{table}
The link distances for the two Tx locations Tx$_\mathrm{a}$ and Tx$_\mathrm{b}$ are $142$~mm and $79$~mm, respectively. The field propagation takes place in different human tissues, such as intestine, muscle, fat, and skin. If setting a unique permittivity along this propagation path to a muscle-fat average, the link distances correspond to $2.4\lambda_{\mathrm{mat}}$ and $1.3\lambda_{\mathrm{mat}}$ at 915~MHz, where $\lambda_{\rm mat}$ is the wavelength in the average tissue. Both distances are smaller than $5\lambda_{\mathrm{mat}}$ far-field boundary considered for electrically small antennas~\cite{Balanis2005},\cite{Stutzman2013}.

The $\mathbf{H}$-matrices were obtained through simulations of 4-port scattering parameters. With the Tx antenna ports labeled as 1 and 2, and the Rx antenna ports as 3 and 4, the matrix $\mathbf{H}$ of the $2\times2$ MIMO link becomes:
\begin{equation}\label{s_and_h}
    \mathbf{H} = 
    \begin{bmatrix}
        S_{31} & S_{32} \\
        S_{41} & S_{42}
    \end{bmatrix}.
\end{equation}
Since the orientation of a capsule antenna inside the human GI tract is random, we simulated the $\mathbf{H}$-matrices for a large number of different orientations of the capsule antenna. This enabled observation of the channel statistics, which are more useful in evaluating link performance than a single instantaneous channel realization. For each link simulation, the Tx antenna orientation was randomized. The rotation angles with respect to the coordinate axes illustrated in Fig.~\ref{fig:simsetup} were drawn from uniform distributions on the interval $[0\degree, \ 360\degree]$ for the $x$ and $z$ axes, and on the interval $[-90\degree, \ 90\degree]$ for the $y$ axis. The rotations were defined in relation to the base orientation, where the capsule was oriented along the $x$ axis as illustrated in Fig.~\ref{fig:simsetup}. Mesh refinements were performed until the magnitude of $\mathbf{H}$-matrix elements changed less than $0.25$~dB between successive refinement runs. To reach this criterion faster, the metallic parts on the capsule and Rx antennas were locally meshed. Simulating $200$ realizations of $\mathbf{H}$ over two link distances with different capsule orientations was sufficient to obtain stable estimates of the 1st, 5th, 25th, 50th, 75th, 95th, and 99th percentile values of the channel capacity~\eqref{eq:capacity} and eigenvalues. The parameters used for the capacity calculation are collated in Table~\ref{tab:capacity_params}.
\begin{table}[h]
    \caption{Parameters utilized in capacity calculation.}
    \centering
    \begin{tabular}{|c|c|}
    \hline
        Parameter & Value \\
       \hline
       Tx Power (dBm)  &  $-43$ \\
       \hline
       Noise Figure (-)/(dB) & $2.5/4$ \\
       \hline
       Temperature (K) & $310$ \\
       \hline
       Bandwidth (MHz) & $1$ \\
       \hline
    \end{tabular}
    \label{tab:capacity_params}
\end{table}
To meet the regulation of $500\ \mu$V electric field at a distance of $3$~m from the body, a transmit power of $-43$~dBm was chosen for the analysis~\cite{Homan2011}. Compared to the $-10$~dBm transmit power in the IEEE 802.15.6 standard~\cite{IEEEWban}, this choice of transmit power offers a larger safety margin to the SAR limit and reduces RF interference with other devices. 

\subsection{Scattering parameters, Channel Capacity and Eigenvalues}
\label{sec:linksim_body_phantom}
\begin{figure}[h]
    \centering
    \includegraphics[width=\columnwidth]{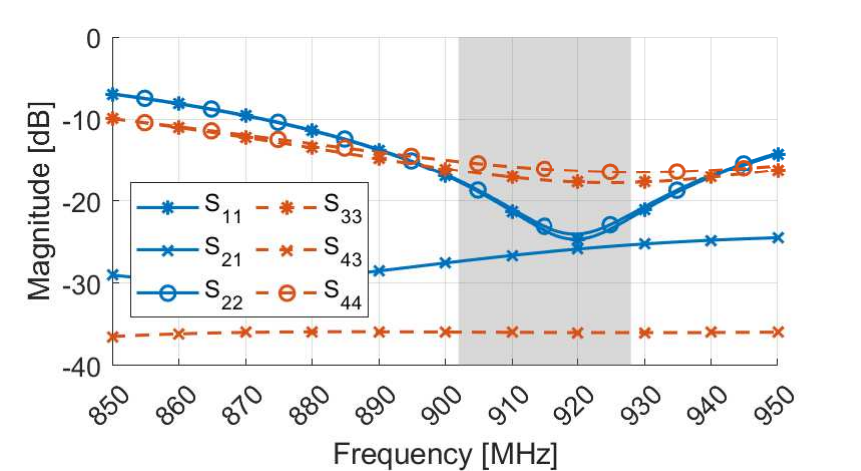}
    \caption{Simulated reflection and coupling scattering parameters corresponding to the setup in Fig.~\ref{fig:simsetup}. Capsule Tx ports use labels 1 and 2 and are plotted in blue, while on-body Rx ports are labeled as 3 and 4 and are plotted in red. The $915$~MHz ISM band is indicated by the gray rectangle.}
    \label{fig:sp_from_sim}
\end{figure}

Figure~\ref{fig:sp_from_sim} plots the simulated reflection and coupling scattering parameters corresponding to the Tx and Rx setup in Fig.~\ref{fig:simsetup}.
The $915$~MHz ISM band is highlighted in the same plot with a gray rectangle. Antennas at both link ends feature reflection coefficients below $-10$~dB and coupling levels below $-20$~dB throughout the band. 
 
\begin{figure}[h]
    \centering
    \includegraphics[width=\columnwidth]{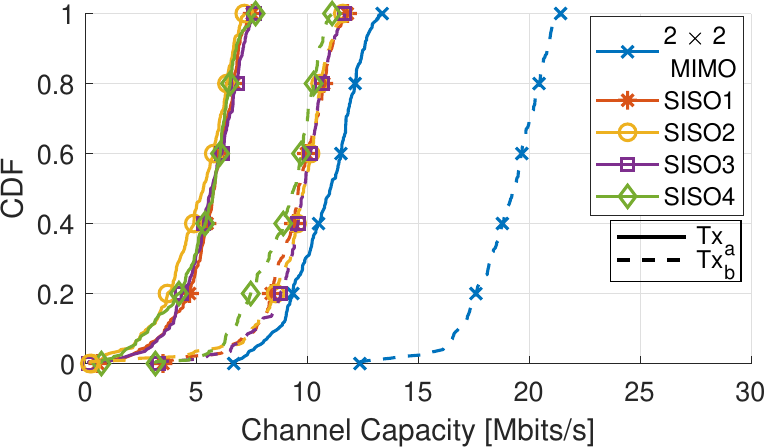}
    \caption{CDF of simulated channel capacity at 915~MHz, 200 links with different orientations of the capsule. Tx$_\mathrm{a}$ and Tx$_\mathrm{b}$ refer to the two Tx locations specified in Table~\ref{tab:positions}.}
    \label{fig:cap_MIMO_vs_SISO}
\end{figure}

The cumulative distribution functions (CDFs) of simulated instantaneous capacities of the $2\times2$ MIMO channel~\eqref{eq:capacity} and the four contained single-antenna channels SISO1$,\, \ldots,$\,SISO4 from the setup of Fig.~\ref{fig:simsetup} are shown in Fig.~\ref{fig:cap_MIMO_vs_SISO}.
The contained SISO channels refer to the four single-Tx single-Rx channels, equivalent to the channel matrix entries in~\eqref{s_and_h}. The obtained MIMO median capacities are $11$~Mbit/s and $19$~Mbit/s for Tx$_\mathrm{a}$ and Tx$_\mathrm{b}$, being $1.90$ and $1.96$ times the best single-antenna median capacities of $5.8$~Mbit/s and $9.8$~Mbit/s. This is because the proposed two-element antenna takes advantage of polarization multiplexing, regardless of the orientation and location in the body.

\begin{figure}[h]
    \centering
    \includegraphics[width=\columnwidth]{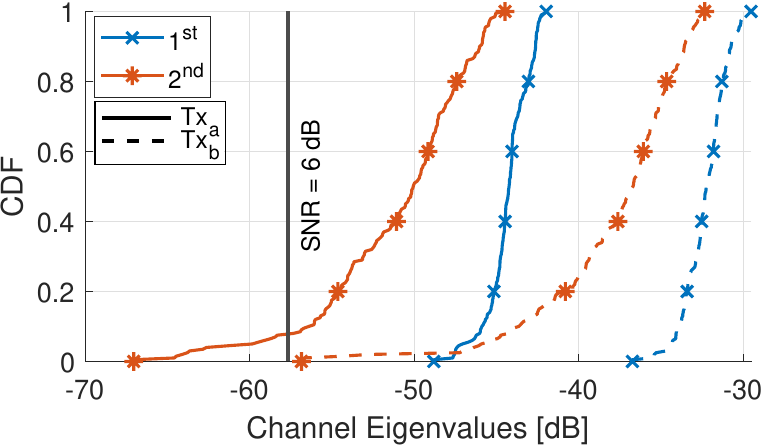}
    \caption{CDF of simulated eigenvalues of the MIMO channel at 915~MHz, 200 links with varying orientations of the capsule. Tx$_\mathrm{a}$ and Tx$_\mathrm{b}$ refer to the two Tx locations specified in Table~\ref{tab:positions}.}
    \label{fig:eigratios}
\end{figure}

To further examine the polarization multiplexing fitness of the proposed antenna, channel eigenvalues were calculated for the $200$ simulated $\mathbf{H}$-matrices. Figure~\ref{fig:eigratios} shows their CDFs.
The vertical line illustrates the $SNR = 6$~dB threshold, below which the channel can be considered rank-deficient and ill-conditioned. For deeper Tx$_\mathrm{a}$, the median eigenvalues are $-44$~dB and $-50$~dB. Correspondingly, they are $-32$~dB and $-37$~dB for shallower Tx$_\mathrm{b}$. The larger median values for Tx$_\mathrm{b}$ are explained by a greater median SNR of $29$~dB for Tx$_\mathrm{b}$, compared to a median SNR of $17$~dB for Tx$_\mathrm{a}$. All the realized links for Tx$_\mathrm{b}$ were full-rank, while for Tx$_\mathrm{a}$, $7$~\% of the channel was rank-deficient given the conservatively low $-43$~dBm total transmit power. In that case, channel state information at the Tx and the use of water-filling scheme~\cite{Molisch2011, Goldsmith2012} would minimize the penalty of the rank deficiency on capacity, where outage eigenchannels are allocated with almost no power.

\subsection{Comparison with Other Two-Element Antennas}
\begin{figure}[h]
    \centering
    \hspace{-10pt}
    \subfloat[\label{fig:ourant}]{\includegraphics[height=2.2cm]{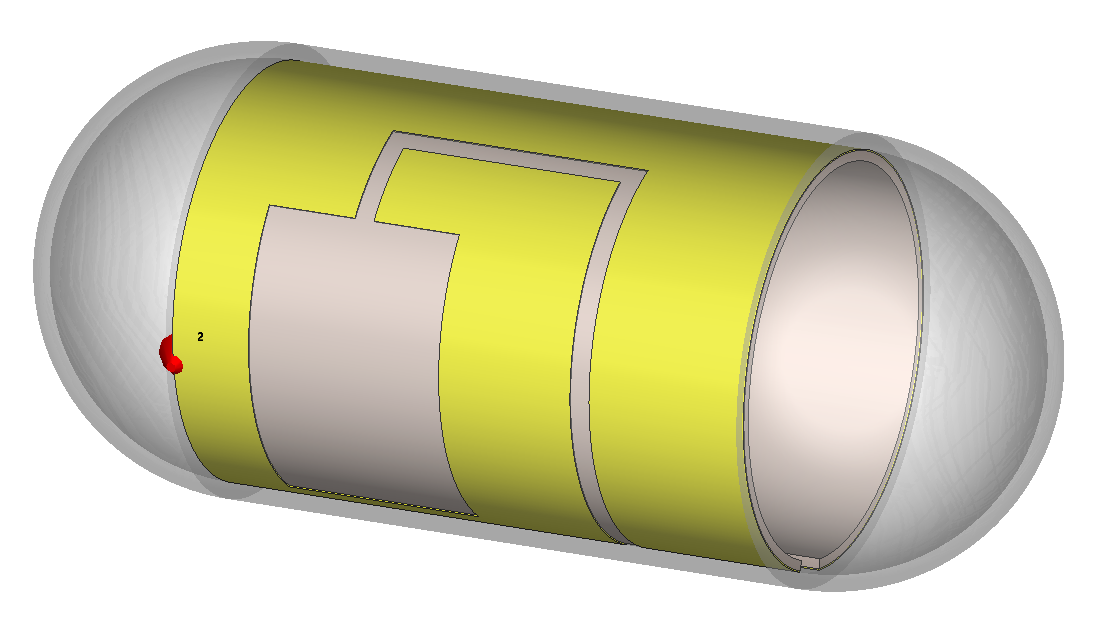}}
    \hspace{30pt}
    \subfloat[\label{fig:sun2025}]{\includegraphics[height=2cm]{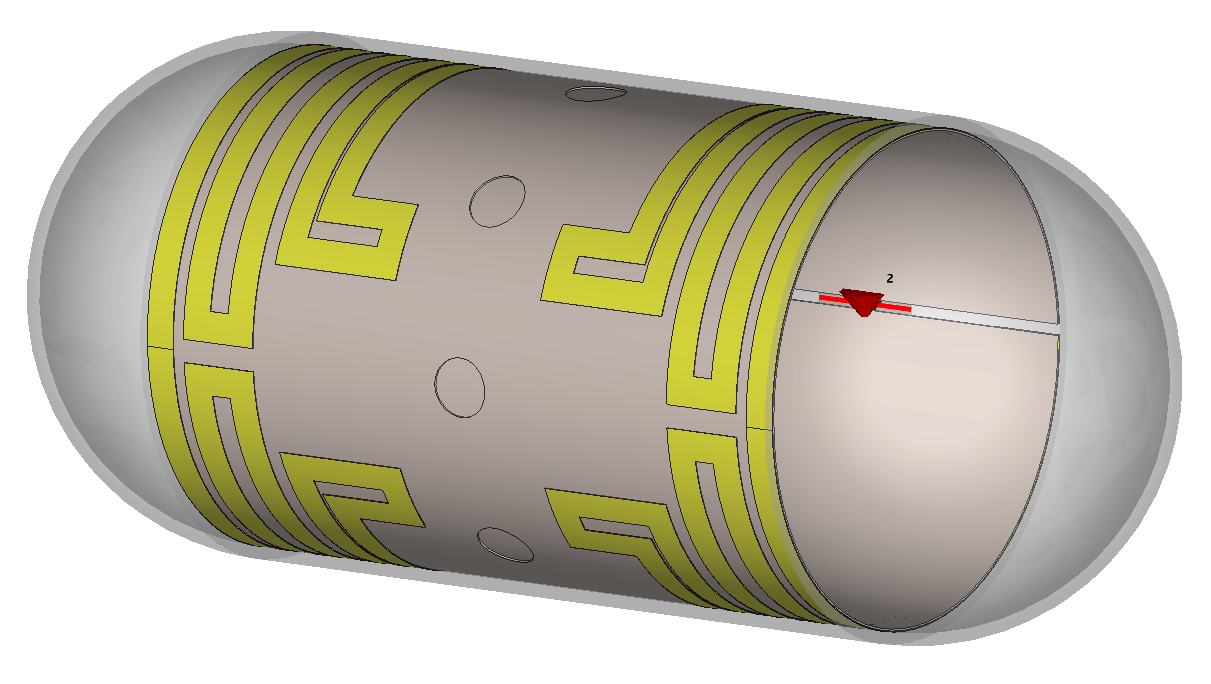}}
    \hfill
    \subfloat[\label{fig:chang2024}]{\includegraphics[height=2cm]{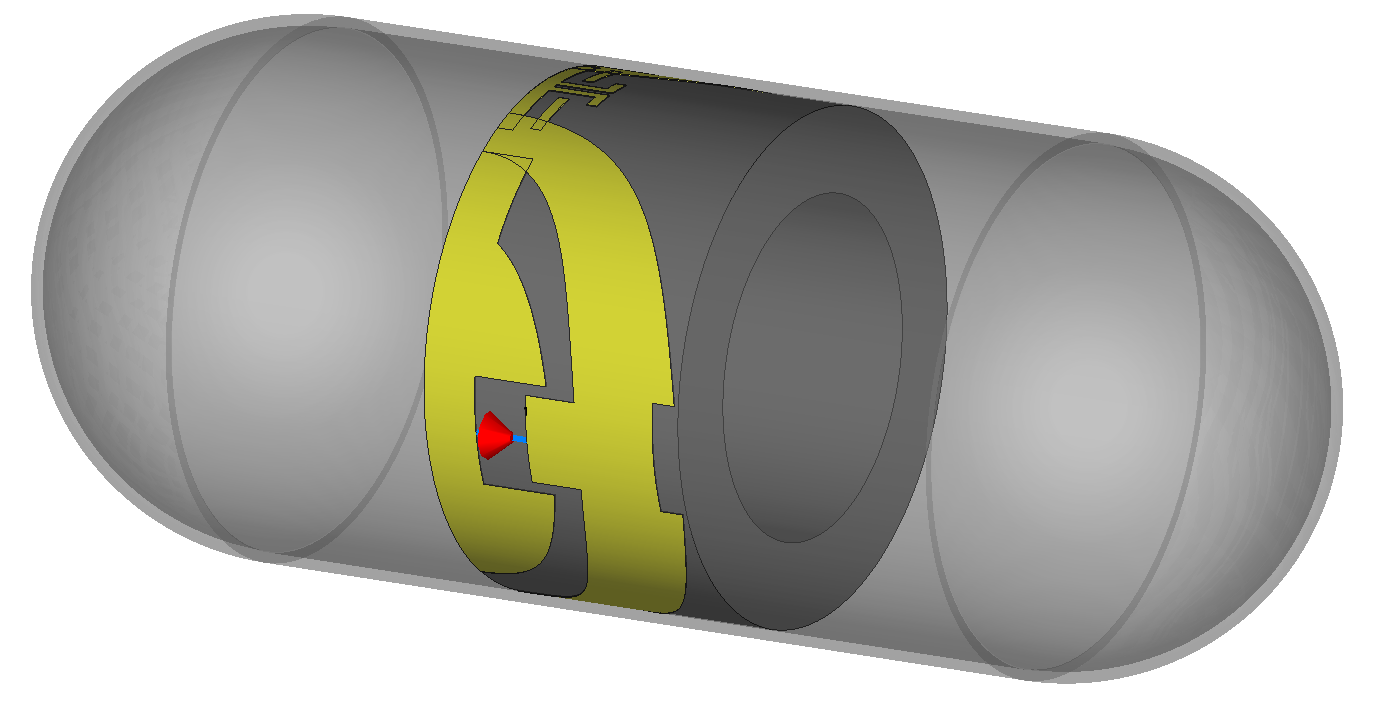}}
    \hspace{20pt}
    \subfloat[\label{fig:li2025}]{\includegraphics[height=2cm]{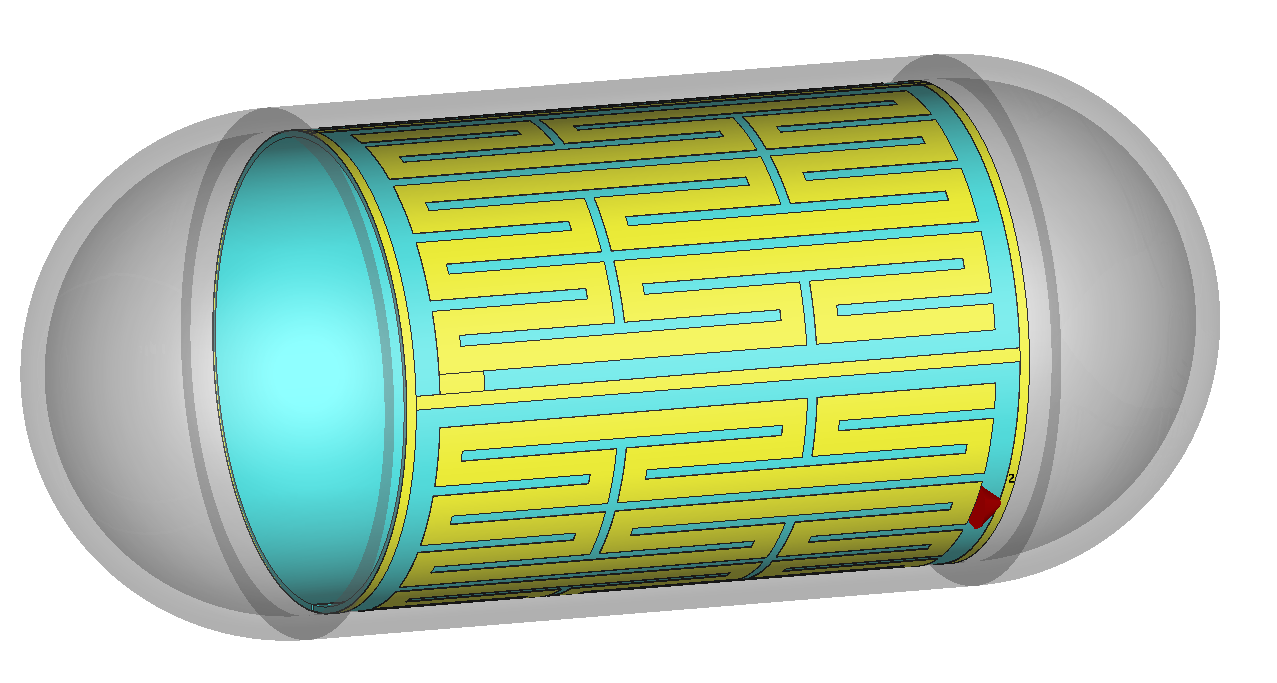}}
    \caption{Illustration of two-element conformal WCE capsule antennas operating at 915 MHz: a) proposed antenna b) \cite{Chang2024} c) \cite{Li2025}, d) \cite{Sun2025}.}
    \label{fig:comparedantennas}
\end{figure}

In this subsection, we benchmark our antenna against other two-element conformal antennas operating at $915$~MHz. Benchmarking was performed by simulating the antennas in the setup of Fig.~\ref{fig:simsetup} and comparing the obtained channel capacities with the same $-43$~dBm total transmit power. The compared antennas are illustrated in Fig.~\ref{fig:comparedantennas}.
The antennas were reconstructed as accurately as possible so that the reported matching and coupling levels in the corresponding literature were reproduced. However, incomplete mentioning of dimensions or material parameters in certain antennas~\cite{Chang2024} and~\cite{Li2025} prevented perfect reproduction of the reported matching levels. Hence, ideal L-section matching circuits were generated for those antennas using OptenniLab~\cite{OptenniLab} circuit synthesis software. With this procedure, all the considered antennas showed matching and coupling of $-9$ dB and $-10$ dB or better at $915$~MHz, ensuring a meaningful comparison. All antennas, except~\cite{Li2025}, were encapsulated in a typical $26$-by-$11$~mm$^2$ standard capsule, while the capsule diameter of the antenna from~\cite{Li2025} was $10$~mm. Link conditions in the comparison follow Table~\ref{tab:capacity_params}, where the Tx and noise power are fixed; $200$ links with varying capsule orientations were simulated for the two Tx locations.

\begin{figure}[h]
    \centering
    \includegraphics[width=\columnwidth]{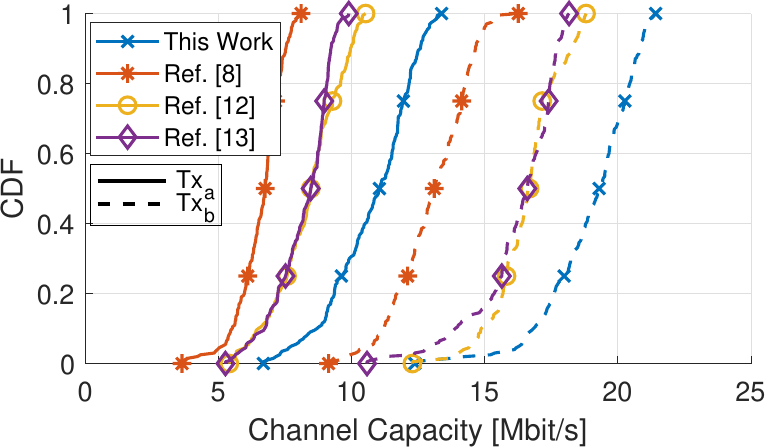}
    \caption{Comparison of the channel capacity CDF for our designed capsule antenna against existing conformal two-element capsule antennas at 915 MHz. The comparison is through simulations of near-field capsule radio links illustrated in Fig.~\ref{fig:simsetup} under the same total transmit power and bandwidth.}
    \label{fig:antennacomp_capacity}
\end{figure}

Figure~\ref{fig:antennacomp_capacity} shows a channel capacity CDF of the antennas, 
demonstrating that our antenna outperforms the other conformal two-element antennas at $915$~MHz across all percentile levels. This result is attributed to its best efficiency and its aligned near-field patterns to the Rx antenna.

\begin{figure}[h]
    \centering
    \includegraphics[width=\columnwidth]{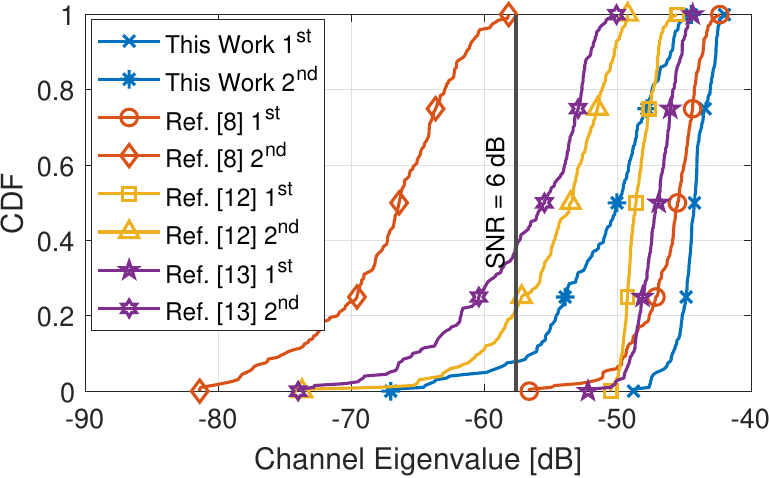}
    \caption{Comparison of the channel eigenvalues for our designed capsule antenna against existing conformal two-element capsule antennas at location Tx$_\mathrm{a}$ at $915$~MHz. The comparison is through simulations of near-field capsule radio links illustrated in Fig.~\ref{fig:simsetup}.}
    \label{fig:antennacomp_eigs}
\end{figure}

The capacity differences across the compared antennas can be further elaborated by observing the statistics of channel eigenvalues $10\log_{10}[\lambda_i(\mathbf{HH}^\mathrm{H})]$, plotted for Tx$_\mathrm{a}$ in Fig.~\ref{fig:antennacomp_eigs}.
This plot shows that the magnitude differences between the first and second eigenvalues for the antennas in~\cite{Li2025} and~\cite{Sun2025} are comparable with those of the proposed antenna. Hence, the multiplexing capabilities of those antennas are similar to the proposed antenna. However, the superiority of the proposed antenna is due to efficiency differences. In the setup of Fig.~\ref{fig:simsetup}, the total efficiency of the proposed antenna is $-38.7$~dB compared to $-41.3$~dB of~\cite{Li2025} and $-41.4$~dB of~\cite{Sun2025}. The higher total efficiency leads to larger eigenvalue magnitudes and a greater percentage of full-ranked channels. On the other hand, the antenna in~\cite{Chang2024} shows the largest difference between the first and second eigenvalues and entirely rank-deficient channels due to a mismatch between Tx and Rx polarizations. This is because the antenna elements in~\cite{Chang2024} have nearly identical polarizations, resulting in a poor polarization pairing with the double-bowtie dipole Rx. In this case, the channel matrices have almost linearly dependent columns, resulting in small second eigenvalues and rank-deficient channels. Conversely, the elements of the proposed antenna have different polarizations in most directions, leading to better polarization matching with the Rx. This results in channel matrices with significantly larger diagonal entries compared to the off-diagonal entries in magnitude, which yields more balanced eigenvalues. The eigenvalue plot for Tx$_\mathrm{b}$ is omitted as it leads to the same conclusions regarding the multiplexing capability and efficiency of the antennas.

\section{Comparison of MIMO Link Simulation and Measurement in Homogeneous Tissue Phantom}\label{sec:MIMOmeas}
Different from discussions so far about the design of our antenna and to show that our antenna outperforms existing designs, the present section focuses on a comparison between simulated and measured characteristics of MIMO channels formed by our capsule antenna, aiming at experimental verification of the antenna design. Since the use of a heterogeneous "Nelly" human 3D model makes a comparison with experiments difficult, we instead perform the comparison with a homogeneous liquid muscle tissue phantom.

\subsection{Homogeneous Liquid Phantom Model}
\label{sec:phantomodel}
\begin{figure}
    \centering
    \includegraphics[width=0.8\columnwidth]{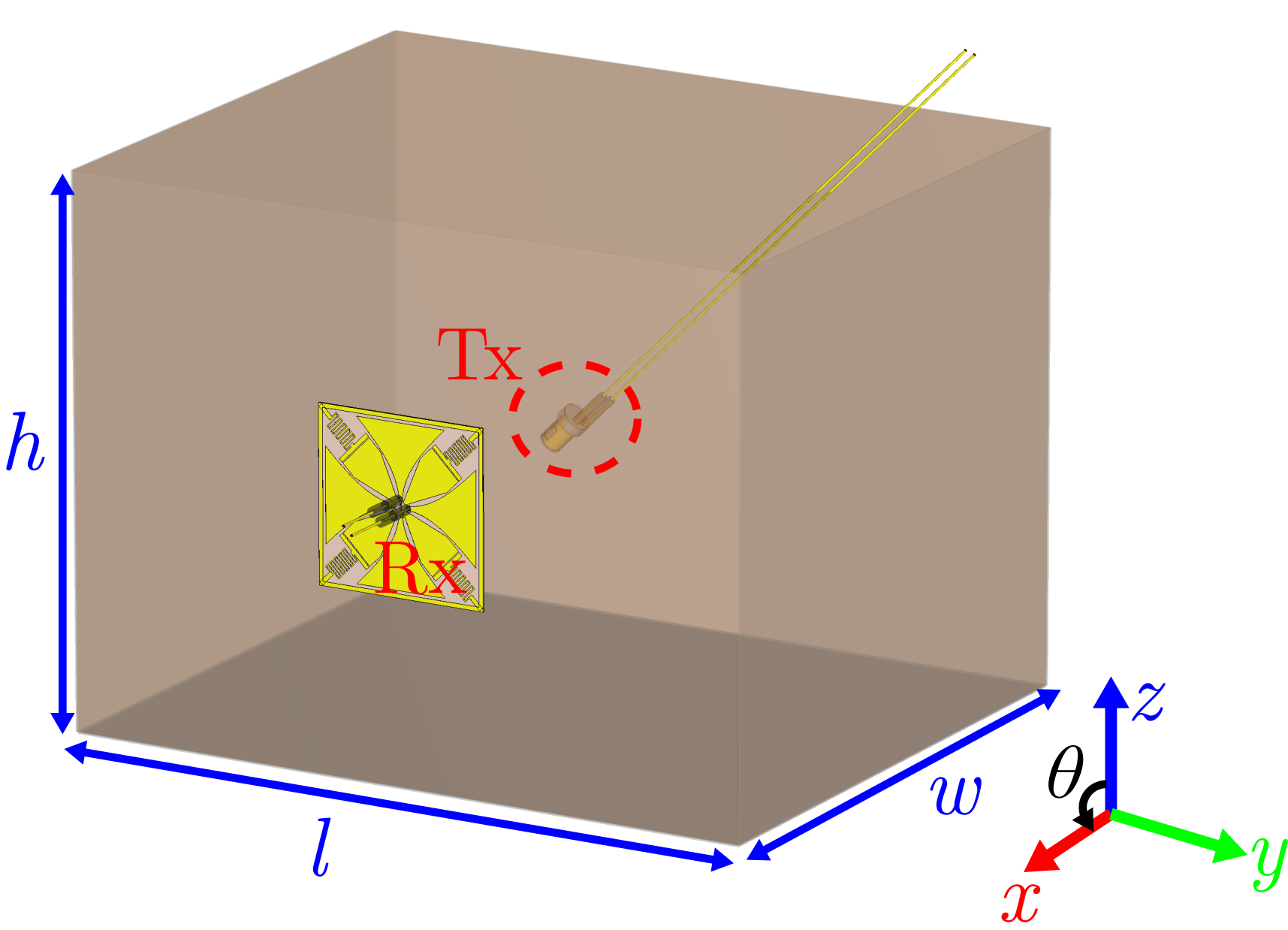}
    \caption{Simulation setup with homogeneous muscle tissue phantom. The phantom dimensions are $l = 340$~mm, $w = 280$~mm, and $h = 260$~mm. The origin of the coordinate system is at the center of the phantom block.}
    \label{fig:meas_simsetup}
\end{figure}

A numerical model of a homogeneous muscle tissue phantom is shown in Fig.~\ref{fig:meas_simsetup}. The model approximates the measurement setup presented in Section~\ref{sec:realmodel}.
The electrical properties of the muscle tissue phantom can be found in~\cite{Kiviharju2025}.  Similarly to the simulations with the "Nelly" model, two locations,  Tx$_\mathrm{c}$, and Tx$_\mathrm{d}$, were simulated. With the coordinate system origin at the center of the phantom in the $x$, $y$, and $z$ directions, the locations of Tx and Rx antennas are given in Table~\ref{tab:positions_meas}.
\begin{table}[th]
    \caption{Positioning of the Tx and Rx antennas}
    \centering
    \begin{tabular}{|c|c|c|c|}
    \hline
        & $x$ (mm) & $y$ (mm) & $z$ (mm) \\
            \hline
       Tx$_\mathrm{c}$ & $0$ & $0$ & $0$ \\
        \hline
       Tx$_\mathrm{d}$ & $90$ & $0$ & $0$ \\
        \hline
         Rx & $140$ & $0$ & $0$ \\
         \hline
\end{tabular}
    \label{tab:positions_meas}
\end{table}
The resulting Tx-Rx distances are $140$~mm for Tx$_\mathrm{c}$ and $50$~mm for Tx$_\mathrm{d}$. In the muscle phantom material, these distances correspond to $3.2\lambda_{\mathrm{mat}}$ and $1.1\lambda_{\mathrm{mat}}$, respectively, meaning that in both locations the link distances were shorter than the $5\lambda_{\mathrm{mat}}$ far-field distance for electrically small antennas~\cite{Balanis2005},\cite{Stutzman2013}. Due to the positional uncertainty of the capsule antenna in the corresponding measurement, a random displacement in the $x$, $y$, and $z$ directions was introduced in the simulations. The displacements were drawn randomly from a zero-mean normal distribution with a standard deviation of $3.5$~mm.

\subsection{Capsule Antenna Prototype}
\begin{figure}[h]
    \centering
     \subfloat[\label{fig:proto_top}]{\includegraphics[height=6cm]{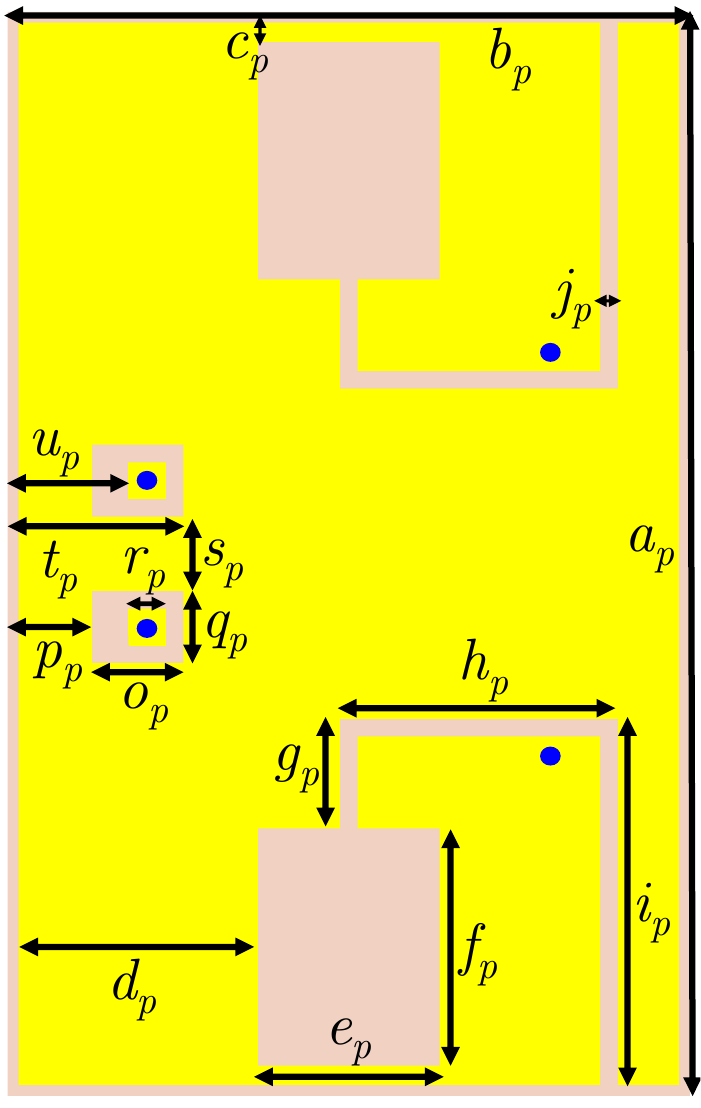}}
     \hspace{10pt}
    \subfloat[\label{fig:proto_bottom}]{\includegraphics[height=6cm]{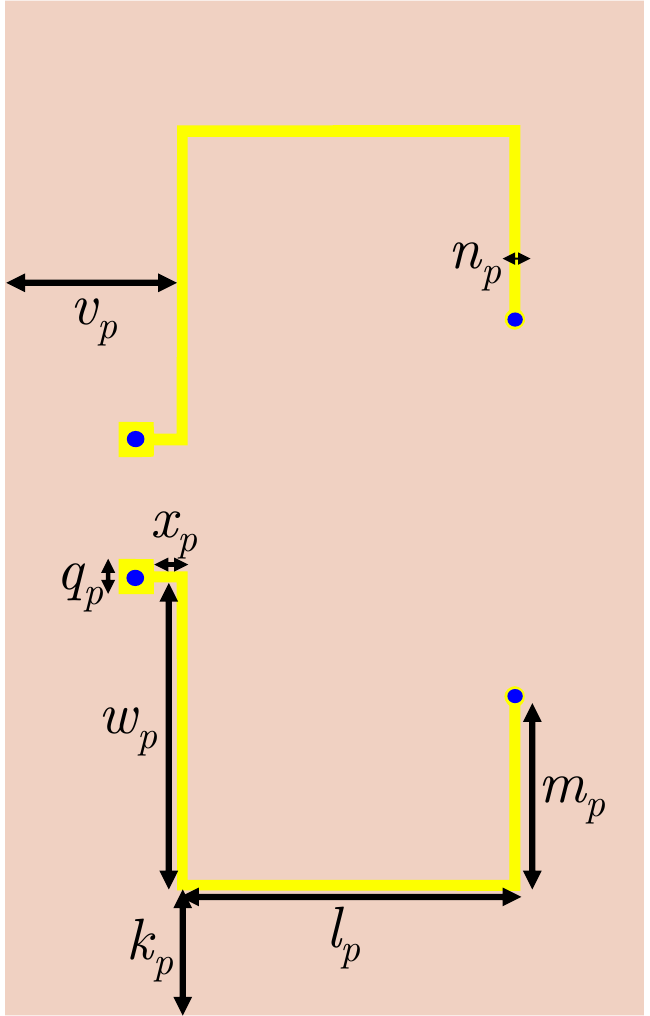}}
    \caption{PCB layout of the fabricated in-body prototype antenna. The yellow color represents copper parts, and the light pink color represents the substrate. Vias are marked in the figures with blue dots. a) Front/outer side b) back/inner side.}
    \label{fig:proto_inbody}
\end{figure}

With the use of a vector network analyzer (VNA), coaxial cables had to be attached to the prototype antenna. This entailed certain modifications to the proposed antenna presented in Section~\ref{sec:antdesign}. These modifications resulted in the prototype PCB shown in Fig.~\ref{fig:proto_inbody}.
The coaxial cables were attached to the front side of the PCB, and a via between the front and back metal layers was added to connect the coaxial center pin to the feeding microstrip on the back side of the PCB. The front and back sides of the antenna PCB became outer and inner sides once conformed into a cylindrical shape. Due to the area occupied by the coaxial connections, the dimension $b$ in Fig.~\ref{fig:ant_top} had to be increased from $15$ to $19$~mm.  The dimensions of the prototype are collated in Table~\ref{tab:prototypedim}.
\begin{table}[h]
    \centering
    \caption{Dimensions of the prototype antenna PCB (mm)}
    \begin{tabular}{|cc|cc|cc|cc|cc|cc|}
    \hline
       a$_p$ & 29.6 & f$_p$  & 6.5 & k$_p$  & 3.7 & p$_p$  & 2.3& u$_p$  & 3.3\\
       \hline
        b$_p$  & 18.6 & g$_p$  & 3.0 & l$_p$  & 10.0 & q$_p$  & 2.0 & v$_p$  & 5.0\\
        \hline
        c$_p$ & 0.8 & h$_p$  & 7.6 & m$_p$  & 5.4 & r$_p$  & 1.0 & w$_p$  & 9.0\\
        \hline
        d$_p$  & 6.8 & i$_p$  & 10.3 & n$_p$  & 0.3 & s$_p$  & 2.0 &  x$_p$ & 1.0\\
        \hline
        e$_p$  & 5.0 & j$_p$  & 0.5 & o$_p$  & 2.5 & t$_p$  & 4.8 & $-$ & $-$\\
        \hline
    \end{tabular}
    \label{tab:prototypedim}
\end{table}
The dimensions are labeled with letters in correspondence to the proposed antenna, distinguished with the subscript $p$ denoting prototype. 

\begin{figure}[h]
    \centering
    \subfloat[\label{fig:proto_unwrapped}]{%
      \includegraphics[width=0.48\columnwidth]{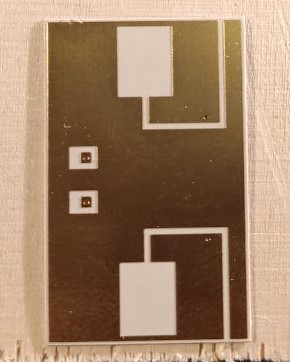} 
    }
    \hfill
    \begin{minipage}{0.48\columnwidth}
        \centering
        \vspace{-4.7cm}
        \subfloat[\label{fig:proto_wrapped}]{%
          \includegraphics[width=0.7\columnwidth]{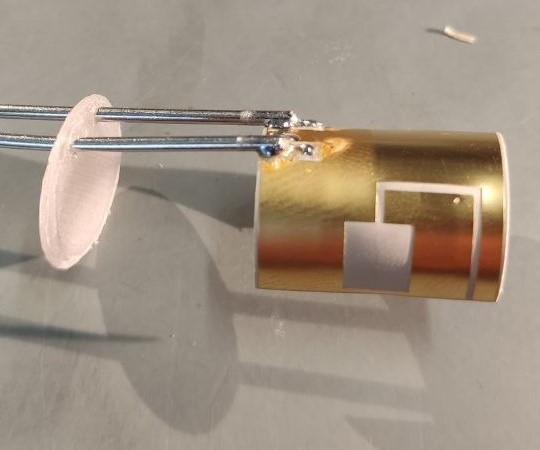} 
        }\\ 
        \vspace{-10pt}
        \subfloat[\label{fig:empty_capsule}]{%
          \includegraphics[width=0.7\columnwidth]{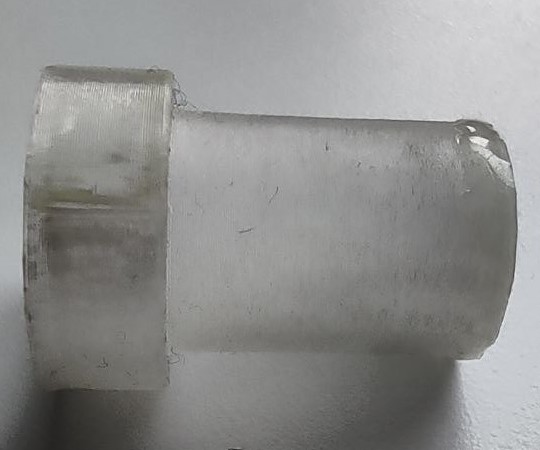}
        }
    \end{minipage}
    \caption{Photograph of the manufactured prototype a) Before conforming b) After conforming and soldering measurement cables c) Empty 3D-printed capsule shell.}
    \label{fig:inbody_photo}
\end{figure}

The prototype antenna was manufactured on a $0.25$~mm thick Rogers RO3010 substrate, with a copper thickness of 18~$\mu {\rm m}$. Figures~\ref{fig:proto_unwrapped} and~\ref{fig:proto_wrapped} show the antenna and its cylindrically conformed shape with feed cables soldered.

\begin{figure}[h]
    \centering
     \subfloat[\label{fig:proto_60}]{\includegraphics[height=5cm]{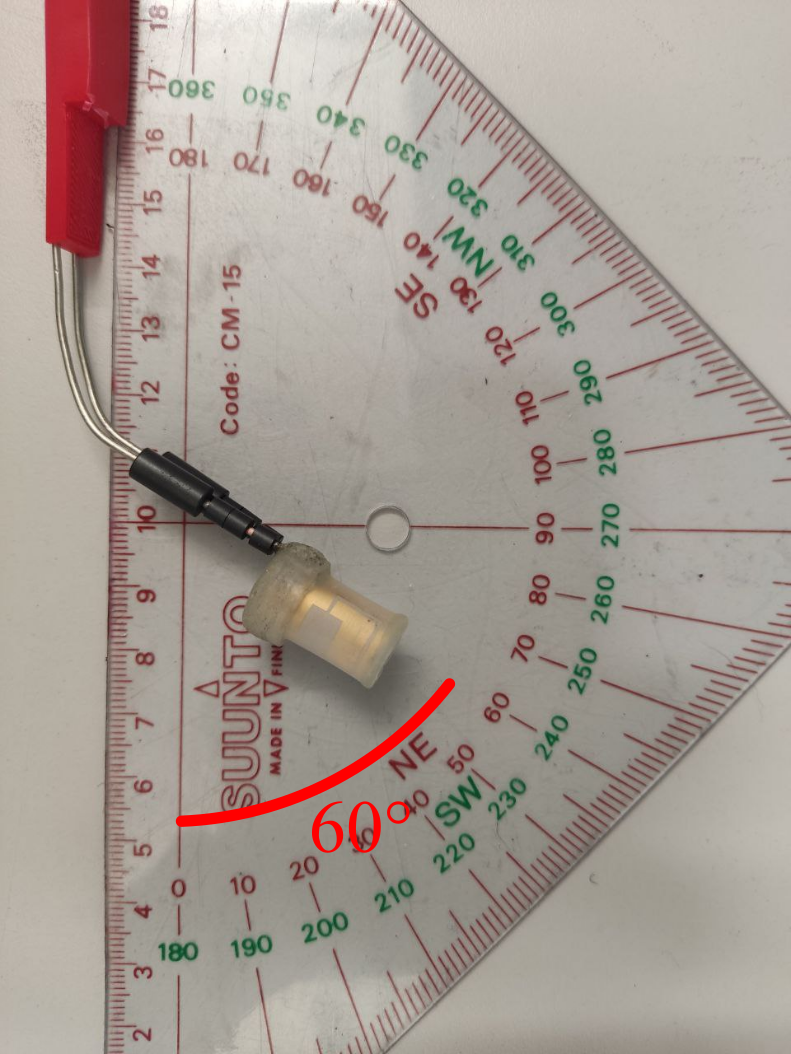}}
     \hspace{10pt}
    \subfloat[\label{fig:proto_120}]{\includegraphics[height=5cm]{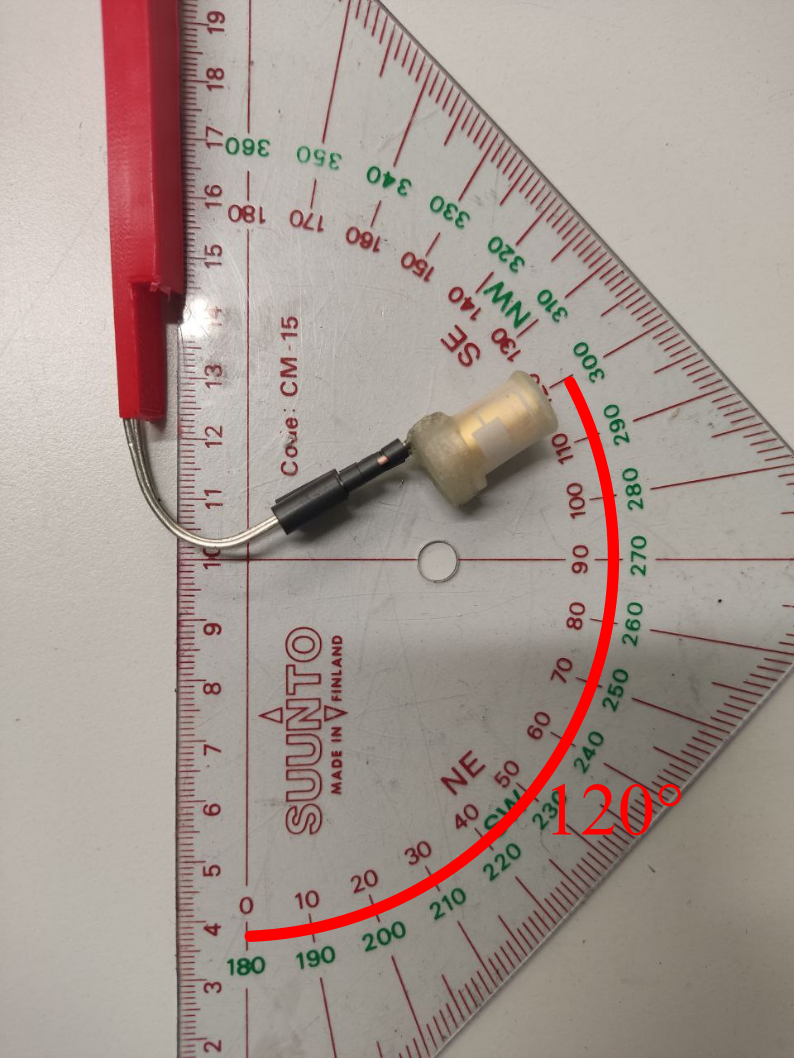}}
    \caption{Illustration of the prototype with cables bent a) 60$\degree$ b) 120$\degree$.}
    \label{fig:proto_bent}
\end{figure}
To prevent direct contact with the phantom liquid, the conformed antenna PCB was inserted into a 3D-printed PLA encapsulation, finally having a wall thickness of approximately 0.3 mm. The encapsulation is shown in Figs.~\ref{fig:empty_capsule} and~\ref{fig:proto_bent}. A cylindrical polystyrene plug was manufactured and placed inside the conformed and encapsulated antenna. This was to minimize the air gap between the antenna PCB and the inner surface of the encapsulation and to ensure a smooth conformation. In addition, the coaxial cables were mechanically reinforced with a support structure shown as the red part in Fig.~\ref{fig:proto_bent} to counteract the buoyant force of the phantom liquid that may otherwise bend the cables.

\subsection{MIMO Link Measurement Setup and Procedure}
\label{sec:realmodel}
\begin{figure}[h]
    \centering
    \includegraphics[width=0.9\columnwidth]{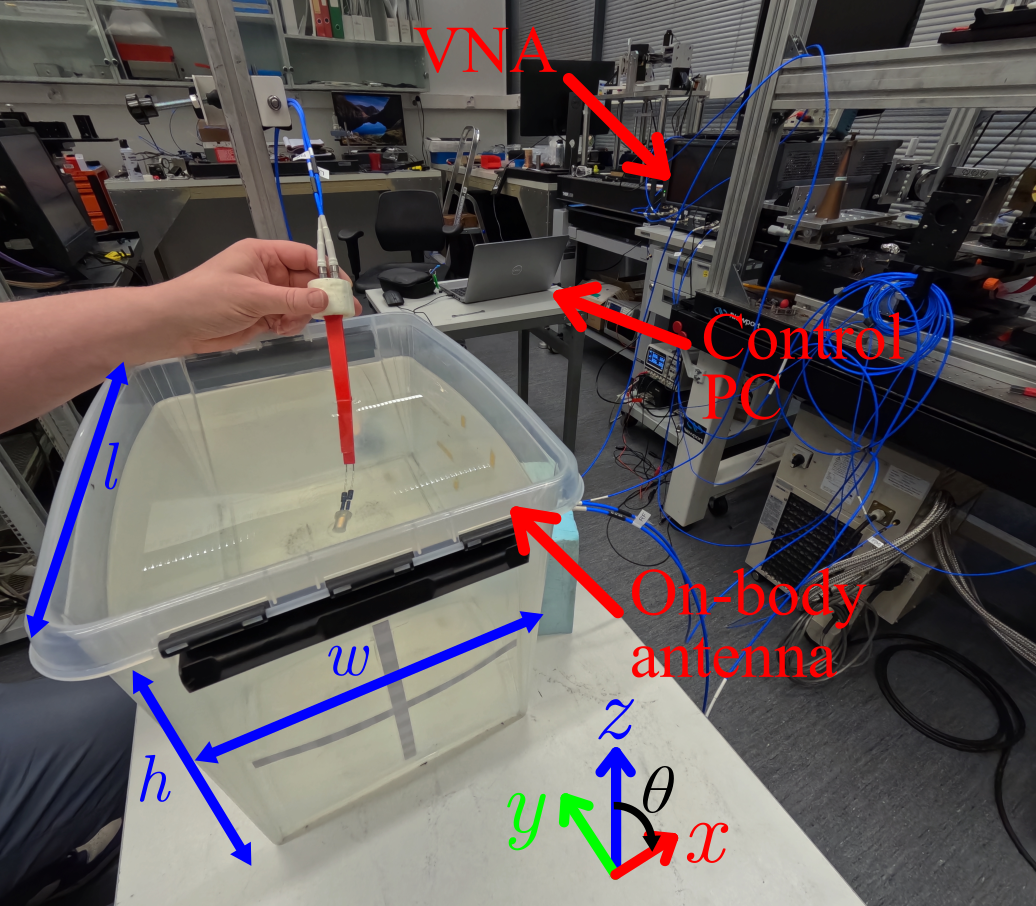}
    \caption{Photograph of the measurement setup. The on-body (Rx) antenna is attached to the side of the phantom box. The phantom dimensions are $l=340$~mm, $w=280$~mm, and $h=260$~mm. The origin of the coordinate system is at the center of the phantom block.}
    \label{fig:meas_setup}
\end{figure}

A four-port VNA (Keysight N5225B) was used to measure~\eqref{s_and_h} from the phantom-immersed Tx antenna ports to the Rx double-bowtie ports located outside the phantom. The measurement setup and its coordinate system are depicted in Fig.~\ref{fig:meas_setup}.
The setup was previously modeled in Fig.~\ref{fig:meas_simsetup} in Section~\ref{sec:phantomodel}. Thus, the Tx and Rx antenna locations were the ones provided in Table~\ref{tab:positions_meas}. A Tx power of $10$~dBm was used, and the intermediate frequency (IF) bandwidth of the VNA was set to $1$~kHz. These settings provided a dynamic range of approximately $80$~dB, which was enough to ensure that none of the measured links were in outage.

Similarly to the simulations with the "Nelly" model and the homogeneous phantom model, the MIMO links were measured with $200$ orientations of the prototype, though the coaxial cables and the supporting structure limited the practically possible orientations. To increase the variation of capsule orientations, three bend angles of the coaxial cables were set, i.e., $0\degree$, $60\degree$, and $120\degree$. The capsules at the latter two bend angles are shown in Fig.~\ref{fig:proto_bent}. The number of measured links was $110$, $70$, and $20$ with bend angles of $0\degree$, $60\degree$, and $120\degree$. In each measurement, the prototype was kept fixed at location Tx$_\mathrm{c}$ or Tx$_\mathrm{d}$, irrespective of the bend angle.

Two operators conducted the measurement of channel matrices as follows. The first operator placed the prototype at the center of the phantom box along the $y$ axis, with angles relative to the $x$, $y$, and $z$ axes being random. Fig.~\ref{fig:meas_setup} can be referred to for the coordinate system. The second operator then ensured the correct Tx position in the $x$ direction and confirmed the centrality in the $z$ direction using the measurement tapes attached to the side of the phantom box. Once the prototype was in the correct position, the measurement of~\eqref{s_and_h} was saved on a PC. Then, the first operator changed the orientation randomly, and the procedure started over again.

\subsection{Comparison between Simulations and Measurements}
\begin{figure}[h]
    \centering
     \subfloat[\label{fig:sp_meas_sim}]{\includegraphics[height=5cm]{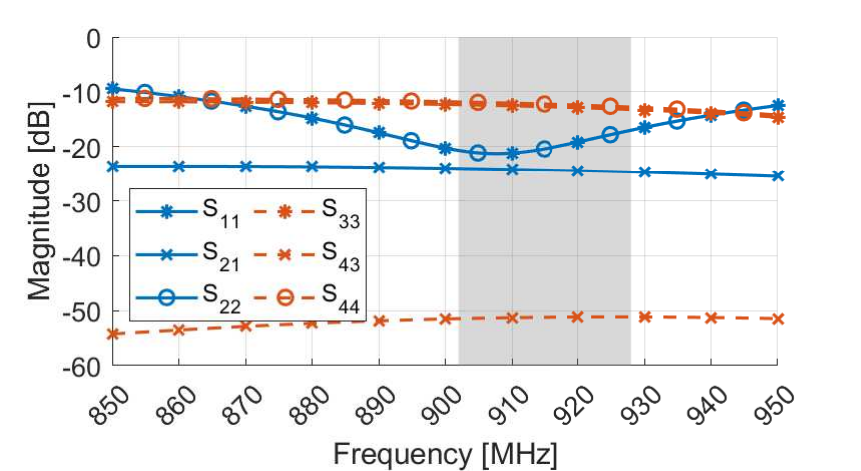}}
     \hspace{10pt}
    \subfloat[\label{fig:sp_meas_meas}]{\includegraphics[height=5cm]{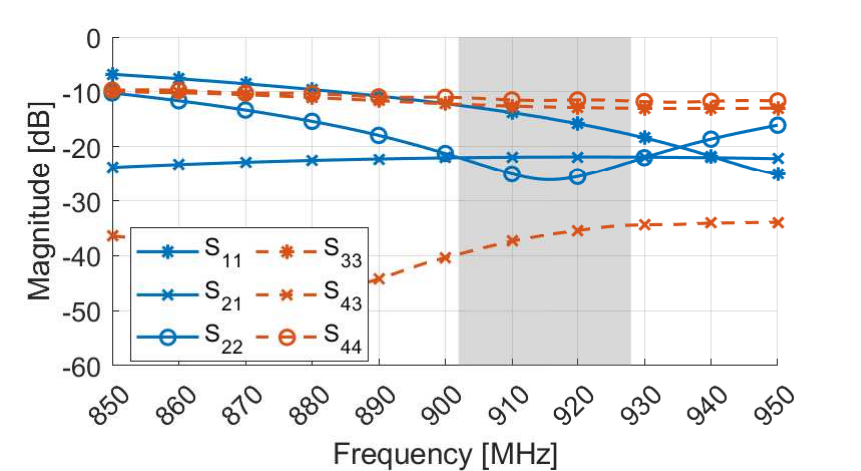}}
    \caption{Scattering parameters of input reflections and mutual coupling for the Tx and Rx prototypes. Capsule Tx ports use labels 1 and 2 and are plotted in blue, while on-body Rx ports are labeled as 3 and 4 and are plotted in red. The $915$~MHz ISM band is indicated by the gray rectangle. a) Simulation b) Measurement.}
    \label{fig:meas_setup_sp}
\end{figure}

Figure~\ref{fig:meas_setup_sp} shows the simulated and measured reflection and coupling scattering parameters for the prototypes of Tx antenna ($S_{11}$, $S_{22}$, and $S_{21}$) and Rx antenna ($S_{33}$, $S_{44}$, and $S_{43}$).
In both simulation and measurement, each antenna element has an input reflection coefficient lower than $-10$~dB, while the inter-element couplings are smaller than $-20$~dB throughout the $915$~MHz ISM band. However, the simulated and measured $S_{11}$ and $S_{22}$ values differ. This is likely caused by the manufacturing tolerance of the Tx antenna prototype. Especially, it is hard to conform the Tx antenna PCB into the encapsulation so that no resonance-affecting air gaps exist between either antenna element and the encapsulation. Nevertheless, as the $-10$~dB matching level is maintained throughout the band of interest, this discrepancy does not affect the link measurement.

\begin{figure}[h]
    \centering
     \subfloat[\label{fig:sim_and_meas_capa_c}]{\includegraphics[height=5cm]{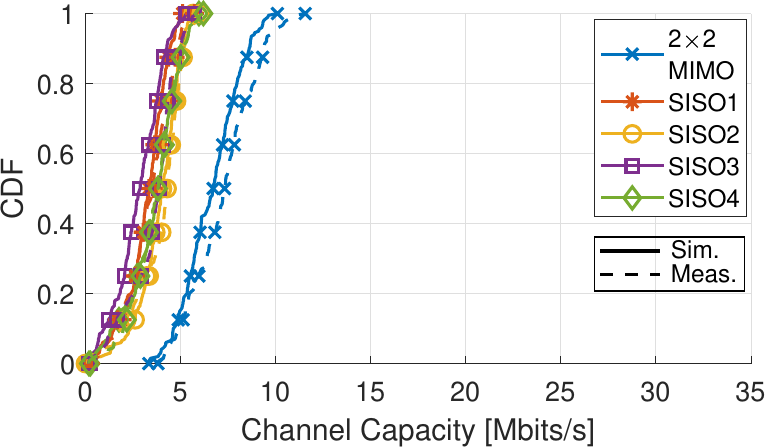}}
     \hspace{10pt}
    \subfloat[]{\includegraphics[height=5cm]{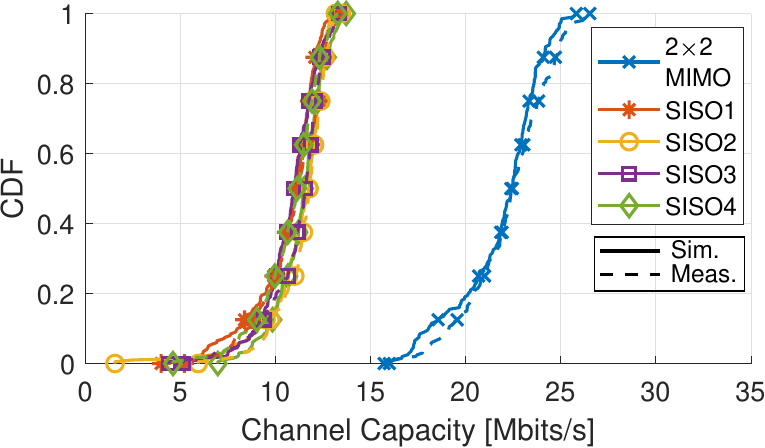}}
    \caption{Capacity CDF of $2\times2$ MIMO and four single-antenna channels at $915$~MHz, obtained from $200$ links with varying capsule orientation. The solid and dashed lines correspond to simulation and measurement: a) Tx$_\mathrm{c}$ and b) Tx$_\mathrm{d}$.}
    \label{fig:sim_and_meas_capa}
\end{figure}

As in Section~\ref{sec:linksim_body_phantom}, channel capacity and eigenvalues were used as metrics for the comparison. The measured channel capacities at $915$~MHz, compared to the FW simulations of the corresponding setup in Fig.~\ref{fig:meas_simsetup}, are provided in Fig.~\ref{fig:sim_and_meas_capa}.
The solid lines correspond to the simulation, and the dashed lines to the measurement. The single-antenna curves correspond to the four possible links of a MIMO link with two Tx and two Rx. The parameters utilized for channel capacity calculation are provided in Table~\ref{tab:capacity_params}. The measured MIMO median capacities are $7.3$~Mbit/s and $22$~Mbit/s for Tx$_\mathrm{c}$ and Tx$_\mathrm{d}$. They are $1.71$ and $1.90$ times those of the best single-antenna median capacities of $4.3$~Mbit/s and $12$~Mbit/s, which is a trend consistent with the body model simulations in Section~\ref{sec:linksim_body_phantom}.

\begin{figure}[h]
    \centering
     \subfloat[]{\includegraphics[height=5cm]{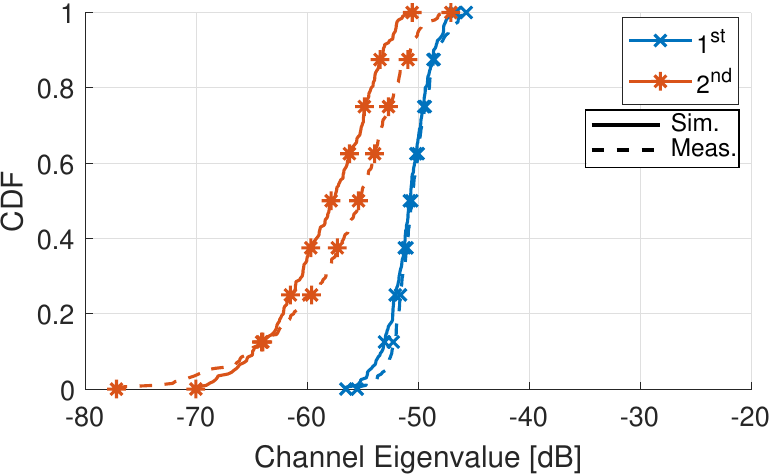}}
     \hspace{10pt}
    \subfloat[]{\includegraphics[height=5cm]{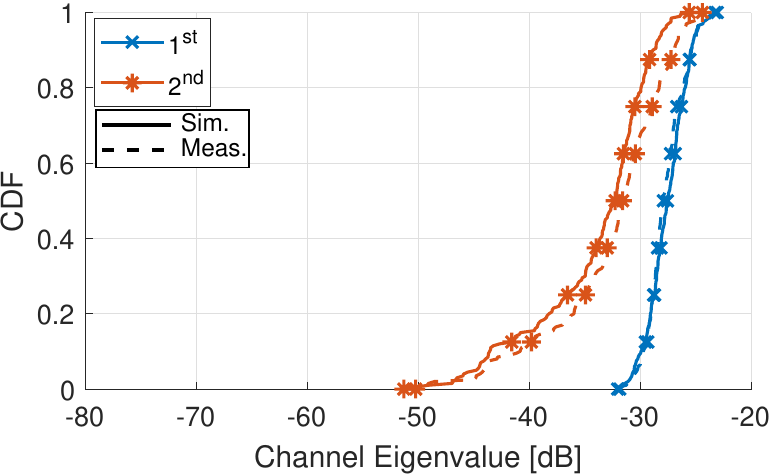}}
    \caption{Channel eigenvalues at $915$~MHz, $200$ links with varying capule prototype orientation. The solid and dashed lines correspond to simulation and measurement, respectively. a) Tx$_\mathrm{c}$ b) Tx$_\mathrm{d}$.}
    \label{fig:sim_and_meas_eigratio}
\end{figure}

Eigenvalues of the simulated and measured channels are plotted in Fig.~\ref{fig:sim_and_meas_eigratio}.
The median values of the first and second measured eigenvalues are $-51$~dB and $-55$~dB for Tx$_\mathrm{c}$, and $-28$~dB and $-32$~dB for Tx$_\mathrm{d}$. The difference is explained by the median SNRs, which are $11$~dB for Tx$_\mathrm{c}$ and $35$~dB for Tx$_\mathrm{d}$. While for both Tx locations, the measured and simulated first eigenvalue agree very well (from the setup of Fig.~\ref{fig:meas_simsetup}), we observe some differences between measured and simulated second eigenvalues. For Tx$_\mathrm{c}$, the lowest simulated eigenvalue was $-70$~dB, while there were $12$ measured channels that gave smaller second eigenvalues than that. This discrepancy can be attributed to the shadowing caused by the measurement cables, which was not modeled exactly in the simulations, because the lowest measured values came from Tx antenna orientations where the cables were partly in the path between the Tx and Rx antennas. Moreover, the median values of the simulated and measured second eigenvalues differ by $2.3$~dB for Tx$_\mathrm{c}$ and $0.7$~dB for Tx$_\mathrm{d}$. The medians of the first and second eigenvalues are closer to each other for the measurements than for the simulations, meaning the polarizations are more orthogonal in the simulation compared to the measurement. A likely reason is the manufacturing tolerance of the Tx antenna prototype. For instance, it is difficult to achieve identical conformation of the antenna PCB in simulation and measurement, causing the near-field patterns of the antenna elements to differ. Furthermore, the slightly larger median value difference of $2.3$~dB for Tx$_\mathrm{c}$ can be explained by the larger link distance and smaller magnitudes of measured channels, i.e., the relative effect of the discrepancies between the measurement setup and the simulation model is larger for smaller measured magnitudes.

The effect of second eigenvalue deviation on channel capacity in Fig.~\ref{fig:sim_and_meas_capa_c} is $9\%$, which is considered low enough to verify the efficacy of the antenna design and link evaluation methods. The proposed antenna can be utilized in polarization multiplexing for improved channel capacity compared to single-antenna counterparts in WCE scenarios.

\section{Conclusion}\label{conclusion}
This article presented a novel two-element capsule antenna for polarization multiplexing at $915$~MHz. The antenna elements featured orthogonal current distributions on a shared aperture, thus enabling polarization multiplexing despite a radio link in the near-field of the antennas. The antenna performance was evaluated with radio link simulations in two environments: the anatomically realistic "Nelly" human body model of CST Studio Suite and a homogeneous muscle tissue phantom. The proposed antenna improved the channel capacity compared to the single-antenna counterpart by a factor of $1.70$ to $1.96$ in both environments and yielded the highest channel capacity when benchmarked against existing conformal two-element WCE capsule antennas at $915$~MHz in the human body model. In addition, the $2\times 2$ MIMO channel with the proposed antenna was full-rank in $97\%$ of the simulated cases in the human body model, averaged over the deep and shallow Tx locations considered. These findings and the agreement of simulated and measured results from the phantom environment together demonstrated that with CSI available at the Tx, the proposed capsule antenna is suitable for polarization multiplexing in the WCE context. As future work, we aim to study a distributed on-body antenna array and include more radiating elements into the capsule so that larger MIMO systems than $2\times2$ are addressed, and to consider other frequency bands designated for medical devices such as $433$~MHz~\cite{Miah19_TAP}.
\bibliography{references.bib}

\end{document}